\documentclass{article}
\usepackage{amsmath}
\usepackage{amssymb}
\usepackage{amsfonts}
\usepackage{bm}
\usepackage[dvips,xdvi]{graphicx}
\usepackage{cite}
\usepackage{color}

\newcommand{\erf}{\operatorname{erf}}

\begin{document}

\noindent{\Large\bf Multi-chain Slip-spring Model for Entangled Polymer Dynamics}\\[10mm]
{Takashi~Uneyama$^{1*}$ and Yuichi~Masubuchi$^{2*}$}\\
{\it $^1$School of Natural System, College of Science and Engineering,
  Kanazawa University, Kakuma, Kanazawa 920-1192,
  Japan}\\
{\it $^2$Institute for Chemical Research, Kyoto University, Gokasho, Uji, 611-0011, Japan}
\\[\baselineskip]
* Corresponding authors\\
e-mail:  uneyama@se.kanazawa-u.ac.jp (TU), mas@scl.kyoto-u.ac.jp (YM)
\\[\baselineskip]
submitted to J. Chem. Phys. on Jul. 4 2012\\
resubmitted on Sep. 13 2012
%
%

\begin{abstract}
 It has been established that the entangled polymer dynamics can be reasonably described by single chain models such as tube and slip-link models. 
 Although the entanglement effect is a result of the hard-core interaction between chains, linkage between the single chain models and the real multi-chain system has not been established yet.
In this study, we propose a multi-chain slip-spring model where bead-spring chains are dispersed in space
 and connected by slip-springs inspired by the single chain slip-spring
 model [A. E. Likhtman, Macromolecules {\bf 38}, 6128 (2005)].
In this model the entanglement effect is replaced by the slip-springs, not by the
 hard-core interaction between beads so that 
 this model is located in the niche between conventional multi-chain simulations and single chain models. 
The set of state variables are the position of beads and the
 connectivity (indices) of the slip-springs between beads.
The dynamics of the system is described by the time evolution equation
 and stochastic transition dynamics for these variables. We propose a
 simple model which is based on the well-defined total free-energy
 and the detailed balance condition.
 The free energy in our model contains a repulsive interaction between beads, which compensate the
 attractive interaction artificially generated by the slip-springs. 
 The explicit expression of linear relaxation modulus is also derived by the linear response theory.
{ We also propose a possible numerical scheme to perform simulations.
Simulations reproduced expected bead number dependence 
  in transitional regime between Rouse and entangled dynamics for the chain structure,
 the central bead diffusion, and the linear relaxation modulus.}
 \\[\baselineskip]
\end{abstract}

\section{Introduction}
It has been rather established that the entangled polymer dynamics can be 
 reasonably described by single chain models where the effect of entanglement
  is replaced by dynamical constraints such as tubes or slip-links\cite{deGennes,Doi,WatanabeRev,McLeish}. 
For instance, several single chain models\cite{MLD,HuaSchieber,DoiTakimoto,LikhtmanMcLeish,Schieber2003,Likhtman2005,Uneyama2011} 
 have been proposed to reproduce viscoelasticity, 
 diffusion, dielectric relaxation, etc, by taking account the relevant relaxation mechanisms such as
 the reptation\cite{deGennes,Doi}, contour length fluctuation\cite{Doi2,Milner} and thermal\cite{Grassley,Tsenoglou,desCloizeaux} and convective constraint releases\cite{Marrucci_JNNFM,Giovanni_JNNFM}.

Despite the successes of these single chain models,
the linkage between the single chain models and the real situation with multi-chains has not been clarified yet.
There have been lots of attempts to extract the parameters for single chain model from multi-chain molecular
 simulations where the entanglement effect is naturally taken into
 account by the hard-core (excluded volume) interaction\cite{Kremer,Ramos,TzoumanekasTheodorou,Harmandaris}. 
Specifically, the primitive path analysis\cite{Everaers,Kroger,Foteinopoulou,Tzoumanekas,Shanbhag} is a realization of the original idea of entanglement,
 and thus, it is promising not only to extract the parameters, also to
 clarify the microscopic picture (or definition) of the entanglement.
Even if some parameters in single chain models are extracted by the primitive path analysis,
there are still some fundamental difficulties.
In most cases, {the mean-field type} single chain description is assumed rather
a priori, and the assumptions employed in the single chain models have not been fully justified yet.
{For instance, mapping of the cross-correlation between different chains}\cite{GaoWeiner,CaoLikhtman,KrogerLuapMuller,Ramirez2007,MasubuchiSukumaran}) {to the single chain 
 models has not been clarified yet.}
Thus a model which needs {fewer} assumptions and
 based on realistic molecular picture is demanding.

A possible approach in the niche between multi-chain and single chain pictures is the multi-chain 
 model where the entanglement is not described by the hard-core interaction but by a coarse-grained
 manner similar to the single-chain models\cite{Twentangle,JCP2001,Kindt,PaddingRev}.
Actually we have developed such a model called the primitive
 chain network (PCN) model (also referred as NAPLES code) where phantom
 chains are bundled by slip-links to form a network in 3D space\cite{JCP2001}. 
 The PCN simulations have been performed for several 
 systems such as linear polymers\cite{JCP2001,JCP2003,MSMSE2004,JNNFM2008}, symmetric and asymmetric stars\cite{MSMSE2004,JCP2011}, comb branch polymer\cite{RheoActa2012}, polymer blends\cite{MSMSE2004},
 and copolymers\cite{JNS2006}, and it has been showed that the PCN simulations can
 reproduce various rheological properties reasonably. Moreover, even
 under large and fast
 deformations\cite{Furuichi2008,JCP2009,Dambal,Furuichi2010,Kushwaha,Yaoita2011,Yaoita2012}
 the PCN simulations show reasonable consistency with experiments.
To locate the PCN model between multi-chain and single chain models, we have reported a comparison with
 single chain models on bidisperse blends\cite{Macro2008} and a comparison with molecular simulations on the network
 statistics\cite{JCP2010}.
However, the total free energy of the system in the PCN model is
not well-defined \cite{UneyamaJCP2011}, because
the equations describing dynamics are not based on the free energy
 nor the detail balance. (This is because the PCN dynamics was modeled
 rather empirically.)
As a result, comparison of some static properties of the PCN model with the other models is essentially difficult.
 
In this study, we newly propose another multi-chain model based on the entanglement picture, which is
 the multi-chain slip-spring model inspired by the single chain slip-spring model proposed by Likhtman
 several years ago\cite{Likhtman2005}.
Differently from the PCN model, we define the total free energy for the new model,
 and we employ dynamics model (a time evolution equation and stochastic
 processes) which satisfies the detailed balance condition. Thus our model
 reproduces the thermal equilibrium which is characterized by the free energy.
In the present paper, we report all the equations to construct the
 model, and also a numerical scheme to implement a simulation code.
Then some preliminary results for such as the chain dimension, the linear 
 viscoelasticity and the center-of-mass diffusion, obtained by the simulations, are reported.

\section{Model}
\subsection{Overview}
\begin{figure}[p]
\newpage
\begin{center}\includegraphics[width=10cm]{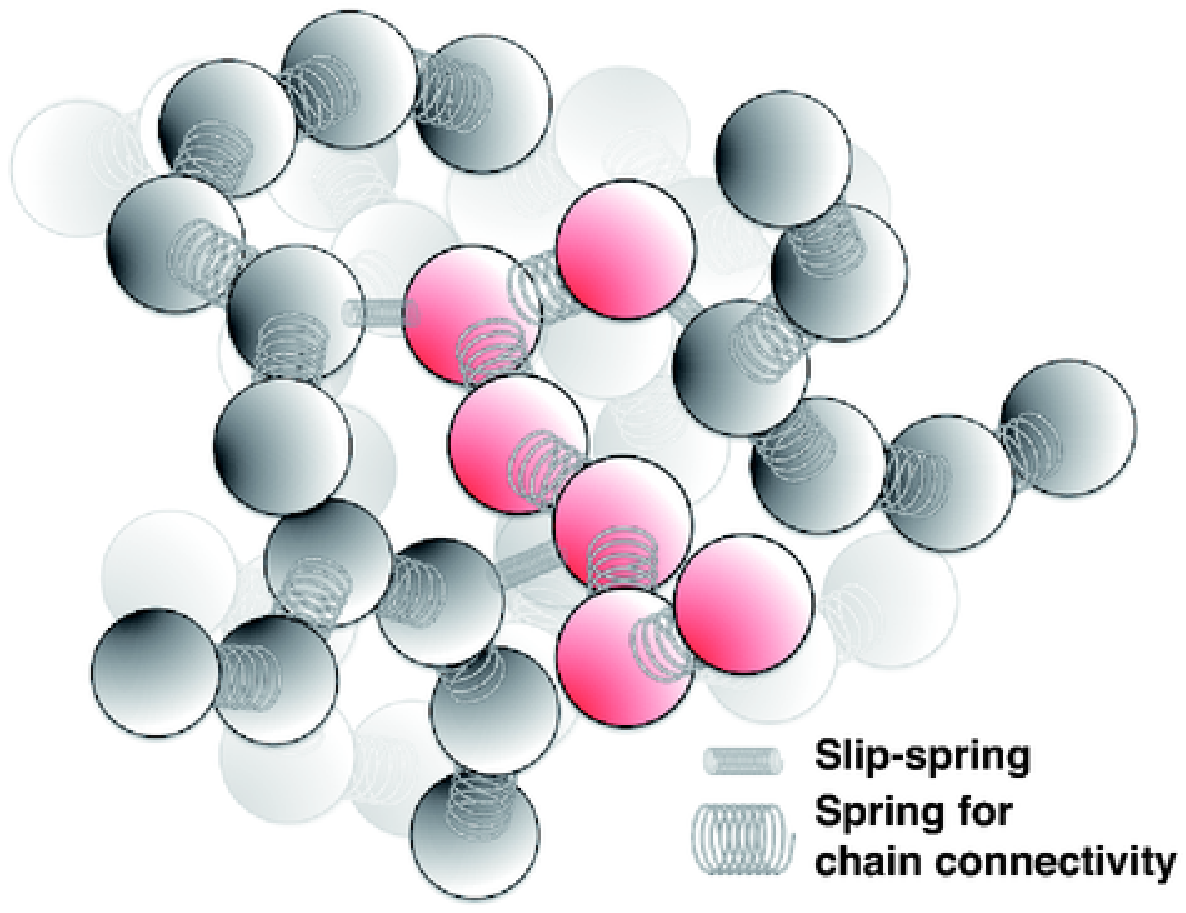}\end{center}
\caption
[ Schematic representation of the model. Bead-spring chains are dispersed in space and connected via slip-springs with each other to form network.]
{T. Uneyama and Y. Masubuchi to {\it J. Chem. Phys}}
\label{Fig:model}
\end{figure}

Figure \ref{Fig:model} shows schematic view of the model employed in this study.
%
%
We consider a network composed of bead-spring chains connected by slip-springs in a volume $V$. 
We describe the number of chains in the system as $M$, and the number of
beads in a chain as $N$.
The beads are connected along the chain backbone by linear entropic springs characterized by the average bond
 size $b$, and chains essentially behave as ideal, Rouse type chains.
Apart from the chain connectivity, some of the beads are connected to the other beads
 by the slip-springs to mimic the entanglement between chains. Following the
 previous works\cite{Likhtman2005, Uneyama2011}, we treat the degrees of freedom of
 slip-spring as state variables, which obey the Maxwell-Boltzmann type
 statistics in equilibrium.
{That is, we assume that the equilibrium probability distribution of
the state variables is described by the Boltzmann weight with the (effective) free energy.}
The ends of slip-springs move (slip) along the chain to reproduce the chain slippage via entanglements.
Instead of the continuous sliding dynamics proposed in the original single-chain model\cite{Likhtman2005}, we
 introduce the discrete hopping dynamics for the end of
 slip-spring between neighboring beads\cite{Uneyama2011}.
The slip-springs are stochastically {destroyed} when one of the ends is at
 the chain ends, and they are also stochastically constructed at the chain ends with a certain probability. 
For the spring force for slip-springs, we employ the linear entropic spring force 
(The strength of this entropic spring is characterized by the slip-spring parameter $N_s$.)
It has been already pointed that slip-links or slip-springs effectively give attractive interaction
between polymers and thus the statistical properties of polymer chains are affected by slip-springs\cite{UneyamaJPSB2011}.
We introduce a repulsive interaction between beads to compensate this attractive
interaction induced by the slip-springs and to recover the ideal chain statistics.
The number density of the slip-spring, $\phi$, is another parameter to
control the strength of the entanglement effect, which is related to the entanglement density.

The state variables of the system are the position of beads, $\{\bm{R}_{i,k}\}$
 (where $i$ and $k$ are the indices for chain and bead position on the chain, respectively),
 the total number of slip-springs in the system $Z$, and the connection matrix of slip-springs $\{\bm{S}_{\alpha}\}$ (where $\alpha$ is  the index for slip-spring and the definition of $\bm{S}_{\alpha}$ is given later).

\subsection{Equilibrium probability distribution}

First we consider the equilibrium statistical properties.
We assume that the chains in our multi-chain slip-spring system obey the
ideal chain statistics. (There is no interaction between beads
except the entropic linear springs.)
Namely,
the probability distribution for the polymer chain conformations
$\lbrace \bm{R}_{i,k} \rbrace$ is given by
\begin{equation}
 \label{equilibrium_probability_gaussian_chains}
 P_{\text{eq}}(\lbrace \bm{R}_{i,k} \rbrace) =
 \frac{1}{V^{M}} \left(\frac{3}{2 \pi b^{2}}\right)^{3 M (N - 1) / 2}
  \exp \left[ - \sum_{i = 1}^{M} \sum_{k = 1}^{N - 1}
	\frac{3 (\bm{R}_{i,k + 1} - \bm{R}_{i,k})^{2}}{2 b^{2}} \right]
\end{equation}
Here the subscript ``eq'' represents the equilibrium quantity.
$V$ is the volume of the system, $b$ is the bead size, $k_{B}$ is the Boltzmann
constant, and $T$ is the temperature.

We consider to put slip-springs in the system, preserving the ideal 
chain statistics mentioned above.
(As we
mentioned, we assume that the system state, including the slip-springs,
is characterized by the free energy.)
For simplicity, we do not introduce restriction for the slip-spring
configurations (connectivity).
For example, we allow multiple slip-springs to share the same bead, or
allow slip-springs to connect two beads on the same chain.
Because both ends of a slip-spring are attached to beads, we need four
indices to specify the state of a slip-spring.
Hereafter, we describe the $k$-th bead on the $i$-th chain as $(i,k)$, and the state
of the $\alpha$-th slip-spring 
as $\bm{S}_{\alpha} \equiv (S_{\alpha,1},S_{\alpha,2},S_{\alpha,3},S_{\alpha,4})$ where
 the ends of the slip-spring are located at the bead $(S_{\alpha,1},S_{\alpha,2})$ and the bead 
 $(S_{\alpha,3},S_{\alpha,4})$.

We assume that there is no specific interaction between
 slip-springs. (This assumption is the same as one employed in single
 chain models, where slip-springs behave as one dimensional ideal
 gas\cite{Likhtman2005,Uneyama2011}.)
The number of slip-springs is not constant and it is controlled by the
 effective chemical potential for slip-springs\cite{SchieberJCP2003}.
Because slip-springs are statistically independent each other, for a given
 polymer conformation, the probability distribution of the slip-spring
 state is given by
\begin{equation}
 \label{equilibrium_probability_slip_spring_position}
 P_{\text{eq}}(\lbrace \bm{S}_{\alpha} \rbrace, Z | \lbrace \bm{R}_{i,k} \rbrace)
  = \frac{1}{\Xi(\lbrace \bm{R}_{i,k} \rbrace)} \frac{1}{Z!}
  \exp \bigg[ - \sum_{\alpha = 1}^{Z} \frac{3 (\bm{R}_{S_{\alpha,1},S_{\alpha,2}} - \bm{R}_{S_{\alpha,3},S_{\alpha,4}})^{2}}{2 N_{s} b^{2}}
   + \frac{\nu Z}{k_{B} T} \bigg]
\end{equation}
where $P(X|Y)$ represents the conditional probability of $X$ under a given
$Y$,
$Z$ is the total number of slip-springs in the system, and $N_{s}$ is the parameter related to the 
 spring constant of slip-spring, and $\nu$ is the effective chemical
 potential for slip-springs. $\Xi(\lbrace \bm{R}_{i,k}
 \rbrace)$ is the grand partition function
  of the slip-spring defined as
\begin{equation}
 \label{grand_partition_function_slip_spring_definition}
 \Xi(\lbrace \bm{R}_{i,k} \rbrace)
  \equiv \sum_{Z = 0}^{\infty} \sum_{\lbrace \bm{S}_{\alpha} \rbrace} \frac{1}{Z!}
  \exp \bigg[ - \sum_{\alpha = 1}^{Z} \frac{3 (\bm{R}_{S_{\alpha,1},S_{\alpha,2}} - \bm{R}_{S_{\alpha,3},S_{\alpha,4}})^{2}}{2 N_{s} b^{2}}
   + \frac{\nu Z}{k_{B} T} \bigg]
\end{equation}
Here $\sum_{\lbrace \bm{S}_{\alpha} \rbrace}$ is taken for all possible
slip-spring indices.
Eq \eqref{grand_partition_function_slip_spring_definition} can be
calculated as follows.
\begin{equation}
 \label{grand_partition_function_slip_spring_modified}
  \begin{split}
   \Xi(\lbrace \bm{R}_{i,k} \rbrace)
   & = \sum_{Z = 0}^{\infty}  \frac{1}{Z!} \bigg[ e^{\nu / k_{B} T} \sum_{\bm{S}}
  \exp \bigg[ - \frac{3
   (\bm{R}_{S_{1},S_{2}} -
   \bm{R}_{S_{3},S_{4}})^{2}}{2 N_{s} b^{2}} \bigg]
   \bigg]^{Z} \\
   & = \exp\bigg[ e^{\nu / k_{B} T} \sum_{\bm{S}}
  \exp \bigg[ - \frac{3
   (\bm{R}_{S_{1},S_{2}} -
   \bm{R}_{S_{3},S_{4}})^{2}}{2 N_{s} b^{2}} \bigg]
   \bigg] \\
   & = \exp\bigg[ e^{\nu / k_{B} T} \sum_{i,k,j,l}
  \exp \bigg[ - \frac{3
   (\bm{R}_{i,k} -
   \bm{R}_{j,l})^{2}}{2 N_{s} b^{2}} \bigg]
   \bigg]
  \end{split}
\end{equation}
Substituting eq \eqref{grand_partition_function_slip_spring_modified} into 
 eq \eqref{equilibrium_probability_slip_spring_position}, we obtain
\begin{equation}
 \label{equilibrium_probability_slip_spring_position_modified}
  \begin{split}
   P_{\text{eq}}(\lbrace \bm{S}_{\alpha} \rbrace, Z | \lbrace \bm{R}_{i,k} \rbrace)
   & = \frac{1}{Z!} \exp\bigg[ - e^{\nu / k_{B} T} \sum_{i,k,j,l}
  \exp \bigg[ - \frac{3
   (\bm{R}_{i,k} -
   \bm{R}_{j,l})^{2}}{2 N_{s} b^{2}}
   \bigg] \\
   & \qquad - \sum_{\alpha = 1}^{Z} \frac{3 (\bm{R}_{S_{\alpha,1},S_{\alpha,2}} - \bm{R}_{S_{\alpha,3},S_{\alpha,4}})^{2}}{2 N_{s} b^{2}}
   + \frac{\nu Z}{k_{B} T} \bigg]   
  \end{split}
\end{equation}
Eqs \eqref{equilibrium_probability_gaussian_chains} and \eqref{equilibrium_probability_slip_spring_position_modified} 
 give the equilibrium distribution function of the full set of state variables as
\begin{equation}
 \label{equilibrium_probability_slip_spring_position_final}
 \begin{split}
  P_{\text{eq}}(\lbrace \bm{R}_{i,k} \rbrace,\lbrace \bm{S}_{\alpha} \rbrace,Z)
  & = 
  \frac{1}{Z!} \frac{1}{V^{M}} \left(\frac{3}{2 \pi b^{2}}\right)^{3 M (N - 1) / 2}
  \exp \bigg[ \frac{\nu Z}{k_{B} T}
  - \sum_{i,k}
	\frac{3 (\bm{R}_{i,k + 1} - \bm{R}_{i,k})^{2}}{2 b^{2}} \\
  & \qquad - \sum_{\alpha = 1}^{Z} \frac{3 (\bm{R}_{S_{\alpha,1},S_{\alpha,2}} -
  \bm{R}_{S_{\alpha,3},S_{\alpha,4}})^{2}}{2 N_{s} b^{2}} \\
  & \qquad - e^{\nu / k_{B} T} \sum_{i,k,j,l} \exp
\left[ - \frac{3 (\bm{R}_{i,k} - \bm{R}_{j,l})^{2}}{2 N_{s} b^{2}}
  \right]
  \bigg]
 \end{split}
\end{equation}

The effective free energy corresponds to 
eq \eqref{equilibrium_probability_slip_spring_position_final} is given by
\begin{equation}
 \label{effective_free_energy_slip_spring_position}
 \begin{split}
  \frac{\mathcal{F}(\lbrace \bm{R}_{i,k} \rbrace,\lbrace \bm{S}_{\alpha}
  \rbrace,Z)}{k_{B} T}
  & = 
  \sum_{i,k} \frac{3 (\bm{R}_{i,k + 1} - \bm{R}_{i,k})^{2}}{2 b^{2}}
  + \sum_{\alpha = 1}^{Z} \frac{3 (\bm{R}_{S_{\alpha,1},S_{\alpha,2}} -
  \bm{R}_{S_{\alpha,3},S_{\alpha,4}})^{2}}{2 N_{s} b^{2}} \\
  & \qquad + e^{\nu / k_{B} T} \sum_{i,k,j,l} \exp
\left[ - \frac{3 (\bm{R}_{i,k} - \bm{R}_{j,l})^{2}}{2 N_{s} b^{2}} \right]
 \end{split}
\end{equation}
The free energy \eqref{effective_free_energy_slip_spring_position}
consists of several contributions. The first and second terms in the right hand
side of eq \eqref{effective_free_energy_slip_spring_position} are the
elastic energies of polymer chains and slip-springs,
respectively. The third term in the right hand side of eq
\eqref{effective_free_energy_slip_spring_position} represents
the repulsive interaction which compensates the attractive interaction
caused by slip-springs\cite{UneyamaJPSB2011}.
The repulsive potential is a soft-core Gaussian type and similar to 
 the Flory-Krigbaum potential\cite{Flory}. 
However, it should be emphasized that this repulsive interaction is not introduced to reproduce  
 the excluded volume effect and the overlapping among the chains, but to compensate the artificial
 attraction by slip-springs. Indeed, the Gaussian form comes from the harmonic
 potential of slip-springs.
 This repulsive interaction acts not only for the beads connected to 
 slip-springs but also for the free beads. 
Using the effective free energy given by \eqref{effective_free_energy_slip_spring_position},
eq \eqref{equilibrium_probability_slip_spring_position_final} can be rewritten as
\begin{equation}
 \label{equilibrium_probability_slip_spring_position_final_modified}
 \begin{split}
  P_{\text{eq}}(\lbrace \bm{R}_{i,k} \rbrace,\lbrace \bm{S}_{\alpha} \rbrace,Z)
  & = 
  \frac{1}{Z!} \frac{1}{V^{M}} \left(\frac{3}{2 \pi b^{2}}\right)^{3 M (N - 1) / 2}
  \exp \bigg[ \frac{\nu Z}{k_{B} T} - \frac{\mathcal{F}(\lbrace \bm{R}_{i,k} \rbrace,\lbrace \bm{S}_{\alpha} \rbrace,Z)}{k_{B} T}
\bigg]
 \end{split}
\end{equation}
The equilibrium statistical average, which we describe as $\langle \dots
\rangle_{\text{eq}}$, can be defined as
\begin{equation}
 \langle \dotsb \rangle_{\text{eq}} \equiv \int d\lbrace \bm{R}_{i,k}
  \rbrace \, \sum_{Z = 0}^{\infty}
  \sum_{\lbrace \bm{S}_{\alpha} \rbrace} \dotsb P_{\text{eq}}(\lbrace \bm{R}_{i,k} \rbrace,\lbrace \bm{S}_{\alpha} \rbrace,Z)
\end{equation}

In our model, the number of slip-springs is controlled by the effective
chemical potential $\nu$. However, in practice, the effective chemical
potential is not convenient nor intuitive. Thus, instead of $\nu$, we
 utilize the average number density of slip-springs, $\phi$ as an input parameter.
(In the thermodynamic limit where $M, V \to \infty$ with fixed $\rho_{0}$, 
the average slip-spring number $\langle Z \rangle_{\text{eq}} = \phi V$
becomes an {extensive} variable while $\phi$ is an {intensive} variable.) The equilibrium
average number density of slip-springs can be calculated from the
equilibrium probability distribution given by \eqref{equilibrium_probability_slip_spring_position_final}.
\begin{equation}
 \label{equilibrium_average_number_of_slip_springs}
 \begin{split}
  \phi & \equiv \langle Z / V \rangle_{\text{eq}} \\
  & = \frac{1}{V} \int d\lbrace \bm{R}_{i,k}
  \rbrace \, \sum_{Z = 0}^{\infty}
  \sum_{\lbrace \bm{S}_{\alpha} \rbrace} Z P_{\text{eq}}(\lbrace
  \bm{R}_{i,k} \rbrace,\lbrace \bm{S}_{\alpha} \rbrace,Z) \\
  & = \frac{e^{\nu / k_{B} T}}{V} \int d\lbrace \bm{R}_{i,k}
  \rbrace \,
  \sum_{i,k,j,l} \exp \bigg[
  - \frac{3 (\bm{R}_{i,k} -
  \bm{R}_{j,l})^{2}}{2 N_{s} b^{2}}
  \bigg] P_{\text{eq}}(\lbrace \bm{R}_{i,k} \rbrace) \\
  & = \frac{e^{\nu / k_{B} T}}{V} \int d\bm{r} \, 
  e^{- 3 \bm{r}^{2} / 2 N_{s} b^{2}}
  \bigg\langle 
  \sum_{i,k,j,l}
   \delta(\bm{R}_{i,k} - \bm{R}_{j,k} - \bm{r})
  \bigg\rangle_{\text{eq}}
 \end{split}
\end{equation}
To evaluate $\phi$, 
the pair-correlation function of beads is required. This pair
correlation function can be calculated straightforwardly, since there
is no correlation between different chains. Thus we have
\begin{equation}
 \label{pair_correlation_function_bead}
 \begin{split}
  \bigg\langle \sum_{i,k,j,l}
   \delta(\bm{R}_{i,k} - \bm{R}_{j,k} - \bm{r})
  \bigg\rangle_{\text{eq}}
  & = M \bigg\langle
  \sum_{k,l}
   \delta(\bm{R}_{i,k} - \bm{R}_{i,l} - \bm{r})
  \bigg\rangle_{\text{eq}}
  + \frac{M (M - 1) N^{2}}{V^{2}} \\
  & = \frac{M}{V} \sum_{k,l}
  \left(\frac{3}{2 \pi |k - l| b^{2}}\right)^{3/2}
  \exp\bigg( - \frac{3 \bm{r}^{2}}{2 |k - l| b^{2}} \bigg)
  + \frac{M (M - 1) N^{2}}{V^{2}}
 \end{split}
\end{equation}
Substituting eq \eqref{pair_correlation_function_bead} into eq
\eqref{equilibrium_average_number_of_slip_springs}, we have
\begin{equation}
 \label{equilibrium_average_number_of_slip_springs_final}
 \phi
  = \rho_{0} e^{\nu / k_{B} T} \bigg[ 1
  + \frac{2}{N} \sum_{k = 1}^{N} \sum_{l = 1}^{k - 1} \bigg(\frac{N_{s}}{
  k - l + N_{s}} \bigg)^{3/2} + \bigg(\rho_{0} - \frac{N}{V}\bigg) \bigg(
  \frac{2 \pi N_{s} b^{2}}{3}
  \bigg)^{3/2} \bigg]
\end{equation}
Here, $\rho_{0} \equiv M N / V$ is the average number density of beads.
Eq \eqref{equilibrium_average_number_of_slip_springs_final} indicates that the
 average slip-spring number depends not only on the effective chemical potential
 but also on the bead density $\rho_{0}$ and the slip-spring intensity $N_{s}$.
This is different from the single chain slip-spring model where the slip-spring 
 density is depends only on the chemical potential $\nu$ \cite{Uneyama2011}.
From eq \eqref{equilibrium_average_number_of_slip_springs_final}, 
 the effective chemical potential $\nu$ corresponding to a given $\phi$ is
 obtained as
\begin{equation}
 \nu
  = - k_{B} T \ln \bigg[ \frac{\rho_{0}}{\phi} \bigg[ 1
  + \frac{2}{N} \sum_{k = 1}^{N} \sum_{l = 1}^{k - 1} \bigg(\frac{N_{s}}{
  k - l + N_{s}} \bigg)^{3/2} + \bigg(\rho_{0} - \frac{N}{V}\bigg) \bigg(
  \frac{2 \pi N_{s} b^{2}}{3}
  \bigg)^{3/2} \bigg] \bigg]
\end{equation}

To calculate rheological properties, we need expression of the
stress tensor of the system. In this work, we employ the following definition for the
stress tensor according to the stress-optical law.
\begin{equation}
 \label{stress_tensor_definition}
 \hat{\bm{\sigma}} \equiv \frac{1}{V} \bigg[ \sum_{i,k} \frac{3 k_{B} T}{b^{2}} (\bm{R}_{i,k + 1} -
  \bm{R}_{i,k}) (\bm{R}_{i,k + 1} - \bm{R}_{i,k}) - M N k_{B} T \bm{1} \bigg]
\end{equation}
Here $\bm{1}$ is unit tensor.
In equilibrium, eq \eqref{stress_tensor_definition} reduces to the
following form.
\begin{equation}
 \begin{split}
  \langle \hat{\bm{\sigma}} \rangle_{\text{eq}}
  & = - \frac{M}{V} k_{B} T \bm{1}
 \end{split}
\end{equation}
 This is the stress tensor of ideal gas, of which number density is $M /
 V$.
The definition of the stress tensor in a slip-spring type model is not
trivial. It is possible to include the contributions from the
slip-springs (contributions from the elastic force and the repulsive force).
Fortunately, most of rheological properties seem to be not sensitive to
the definition of the stress tensor, at least
 qualitatively\cite{Uneyama2011,UneyamaJPSB2011}.
{We will discuss another possible definition of the stress tensor, later.}

\subsection{Dynamics}

While we have specified the equilibrium probability
distribution in the previous section, the dynamical properties such as the
viscoelasticity cannot be determined unless the time evolution equations (rules)
for the state variables are specified. In this section, we design time
evolution equations which satisfy the detailed balance condition. (The
detailed balance condition is required to be satisfied to reproduce the
equilibrium thermodynamical properties correctly.) There are many
possible time evolution equations (rules) which satisfy the detailed balance
condition, and thus we cannot uniquely determine the dynamical model
solely by the equilibrium probability distribution and the detailed
balance condition. In this work, we propose a simple dynamics model which is
similar to the single chain slip-spring models\cite{Likhtman2005,Uneyama2011} and suitable for
numerical simulations.

First, for the time evolution of the bead position $\bm{R}_{i,k}$, we
 employ an overdamped Langevin type equation of motion. In absence of 
the external deformation field, the dynamic equation is given as follows.
\begin{equation}
 \label{langevin_equation_bead}
 \frac{d\bm{R}_{i,k}(t)}{dt}
  = - \frac{1}{\zeta} \frac{\partial \mathcal{F}(\lbrace \bm{R}_{i,k} \rbrace,\lbrace \bm{S}_{\alpha}
  \rbrace,Z)}{\partial \bm{R}_{i,k}} + \bm{\xi}_{i,k}(t)
\end{equation}
Here $\zeta$ is the friction coefficient of a bead and $\bm{\xi}_{i,k}(t)$ is the Gaussian noise
 obeying the fluctuation dissipation relation of the second kind;
\begin{align}
 & \langle \bm{\xi}_{i,k}(t) \rangle = 0 \\
 & \langle \bm{\xi}_{i,k}(t) \bm{\xi}_{j,l}(t') \rangle = \frac{2 k_{B} T}{\zeta}
 \delta_{ij} \delta_{kl} \delta(t - t') \bm{1}
\end{align}
where $\langle \dotsb \rangle$ represents the statistical average.
For some analyses, it would be convenient to introduce the Fokker-Planck
equation.
The Fokker-Plank equation which corresponds to eq \eqref{langevin_equation_bead} is 
\begin{equation}
 \label{fokker_planck_equation_bead}
 \begin{split}
  \frac{\partial P(\lbrace \bm{R}_{i,k} \rbrace,\lbrace \bm{S}_{\alpha}
  \rbrace,Z;t)}{\partial t}
  & = \sum_{i,k} \frac{1}{\zeta} \frac{\partial}{\partial \bm{R}_{i,k}} \cdot
  \left[ \frac{\partial \mathcal{F}(\lbrace \bm{R}_{i,k} \rbrace,\lbrace \bm{S}_{\alpha}
  \rbrace,Z)}{\partial \bm{R}_{i,k}} P
  + k_{B} T \frac{\partial P}{\partial \bm{R}_{i,k}} \right]\\
  & \equiv \mathcal{L}_{\text{FP}} P
 \end{split}
\end{equation}
Here $P(\lbrace \bm{R}_{i,k} \rbrace,\lbrace \bm{S}_{\alpha} \rbrace,Z;t)$
is the time dependent probability distribution and $\mathcal{L}_{\text{FP}}$ is the Fokker-Planck operator.
It is clear that eq \eqref{fokker_planck_equation_bead} satisfies the detailed balance
condition and the steady state distribution coincides to the equilibrium
distribution given by eq \eqref{equilibrium_probability_slip_spring_position_final}.

Second, we consider the reconstruction process of slip-springs.
We assume that the reconstructions of slip-springs are independent of each other, 
and the positions of polymers and other slip-springs do not change
during the reconstruction process.
We write the construction rate of a new slip-spring as $W_{+}(\bm{S}_{Z}, Z | Z - 1)$ and
 the destruction rate of the $\beta$-th slip-spring as $W_{-}(Z - 1|\bm{S}_{\beta},Z)$.
We assume that a slip-spring is {destroyed} with a
 certain fixed probability when one of its ends is at the chain ends.
The destruction rate can be written as
\begin{equation}
 \label{destruction_rate}
 W_{-}(Z - 1| \bm{S}_{\beta}, Z ) 
  = \frac{k_{B} T}{\zeta_{s}}
  \left[ \delta_{S_{Z,2},1} + \delta_{S_{Z,2},N}
   + \delta_{S_{Z,4},1} + \delta_{S_{Z,4},N} \right] \mathcal{E}(\beta,Z)
\end{equation}
Here, $\zeta_{s}$ is the friction coefficient of a slip-spring and 
 $\mathcal{E}(\beta,Z)$ is the exchange operator which exchange the
 $\beta$-th and $Z$-th slip-springs. (By operating
 $\mathcal{E}(\beta,Z)$, $\bm{S}_{\beta}$ and $\bm{S}_{Z}$ are
 exchanged while the other slip-spring indices are unchanged. This operator
 is employed to ensure that the $Z$-th slip-spring is always
 {destroyed}.)

{The slip-spring construction and destruction processes should be
detailed-balanced. The detailed balance condition can be
explicitly written as follows.}
\begin{equation}
 \label{detailed_balance_condition_reconstruction}
  W_{+}(\bm{S}_{Z}, Z | Z - 1) 
  P_{\text{eq}}(\lbrace \bm{R}_{i,k} \rbrace,\lbrace
  \bm{S}_{\alpha} \rbrace,Z - 1)
  =
  \sum_{\beta = 1}^{Z} W_{-}(Z - 1 | \bm{S}_{\beta}, Z) P_{\text{eq}}(\lbrace \bm{R}_{i,k} \rbrace,\lbrace
  \bm{S}_{\alpha} \rbrace,Z)
\end{equation}
{From eqs {\eqref{destruction_rate}} and
{\eqref{detailed_balance_condition_reconstruction}},
the construction rate is uniquely determined.
By substituting eq {\eqref{destruction_rate}} into eq
{\eqref{detailed_balance_condition_reconstruction}},} we obtain the
following explicit form for the construction rate.
\begin{equation}
 \label{construction_rate}
 \begin{split}
  W_{+}(\bm{S}_{Z}, Z | Z - 1) 
  & = \frac{1}{P_{\text{eq}}(\lbrace \bm{R}_{i,k} \rbrace,\lbrace
  \bm{S}_{\alpha} \rbrace,Z - 1)}
  \sum_{\beta = 1}^{Z} W_{-}(Z - 1 | \bm{S}_{\beta}, Z) P_{\text{eq}}(\lbrace \bm{R}_{i,k} \rbrace,\lbrace
  \bm{S}_{\alpha} \rbrace,Z) \\
  & = \frac{k_{B} T}{\zeta_{s}}
  \left[ \delta_{S_{Z,2},1} + \delta_{S_{Z,2},N}
   + \delta_{S_{Z,4},1} + \delta_{S_{Z,4},N} \right] e^{\nu / k_{B} T} \\
  & \qquad \times \exp \bigg[ - \frac{3 (\bm{R}_{S_{Z,1},S_{Z,2}} -
  \bm{R}_{S_{Z,3},S_{Z,4}})^{2}}{2 N_{s} b^{2}}
 \bigg]
 \end{split}
\end{equation}
As before, it would be convenient to introduce the dynamic equation for
the time dependent probability distribution.
The master equation for this reconstruction is written as
\begin{equation}
 \label{master_equation_reconstruction}
 \begin{split}
  \frac{\partial P(\lbrace \bm{R}_{i,k} \rbrace,\lbrace \bm{S}_{\alpha}
  \rbrace,Z;t)}{\partial t}
  & = \sum_{\beta = 1}^{Z + 1} W_{-}(Z | \bm{S}_{\beta}, Z + 1) P(Z + 1)
  + W_{+}(\bm{S}_{Z}, Z | Z - 1) P(Z - 1) \\
  & \qquad - \sum_{\beta = 1}^{Z} W_{-}(Z - 1| \bm{S}_{\beta}, Z)
  P(Z) \\
  & \qquad - \int d\bm{S}_{Z + 1} \, W_{+}(\bm{S}_{Z + 1}, Z + 1 | Z) P(Z)
  \\
  & \equiv \mathcal{L}_{\text{rc}} P
 \end{split} 
\end{equation}
where we have introduced the time evolution operator for the
reconstruction process $\mathcal{L}_{\text{rc}}$.

Finally, we consider the hopping of slip-springs along the chain. 
We assume that there is no interaction between slip-springs and 
 each hopping event is statistically independent.
The hopping process can be described by the change of 
 connectivity index. 
For simplicity, we also assume that the change of connectivity index 
 is restricted as $\pm 1$. (Namely, in our model, the hopping distance of slip-spring on a chain
 corresponds to the bead size).
We consider the event where the $\beta$-th slip-spring changes
its connectivity from $\bm{S}_{\beta}$ to $\bm{S}'_{\beta}$. We
describe the set of slip-spring indices after the hopping as
$\lbrace \bm{S}'_{\alpha} \rbrace$,
for convenience. $\bm{S}'_{\alpha}$ is defined as
\begin{equation}
 \bm{S}_{\alpha}' =
  \begin{cases}
   \bm{S}_{\beta}' & (\alpha = \beta) \\
   \bm{S}_{\alpha} & (\text{otherwise})
  \end{cases}
\end{equation}
%
Let us indicate the transition rate of $\beta$-th slip-spring
 from $\bm{S}_{\beta}$ to $\bm{S}_{\beta}'$ as $W(\bm{S}'_{\beta}|\bm{S}_{\beta})$
 and its transition rate of the inverse process as $W(\bm{S}_{\beta}|\bm{S}'_{\beta})$.
The detailed balance condition can be written as
\begin{equation}
 \label{detail_balanced_condition_jump}
 \begin{split}
  \frac{W(\bm{S}'_{\beta}|\bm{S}_{\beta})}{W(\bm{S}_{\beta}|\bm{S}'_{\beta})}
  & = \frac{P_{\text{eq}}(\lbrace \bm{R}_{i,k} \rbrace,\lbrace \bm{S}_{\alpha}' \rbrace,Z)}
  {P_{\text{eq}}(\lbrace \bm{R}_{i,k} \rbrace,\lbrace \bm{S}_{\alpha}
  \rbrace,Z)} \\
  & =
  \exp \bigg[
 - \frac{3 (\bm{R}_{S_{\beta,1}',S_{\beta,2}'} -
  \bm{R}_{S_{\beta,3}',S_{\beta,4}'})^{2}}{2 N_{s} b^{2}}
  + \frac{3 (\bm{R}_{S_{\beta,1},S_{\beta,2}} -
  \bm{R}_{S_{\beta,3},S_{\beta,4}})^{2}}{2 N_{s} b^{2}} \bigg]
 \end{split}
\end{equation}
If we employ the Glauber type dynamics for the hopping process\cite{Glauber}, the
 transition rate which satisfies eq
 \eqref{detail_balanced_condition_jump} can be expressed as follows.
\begin{equation}
 \label{hopping_rate}
 \begin{split}
 W(\bm{S}_{\beta}'|\bm{S}_{\beta})
  & = \frac{k_{B} T}{\zeta_{s}} \delta_{S_{\beta,1}, S_{\beta,1}'}
  \delta_{S_{\beta,3}, S_{\beta,3}'}
  \bigg[ \left(\delta_{S_{\beta,2}, S_{\beta,2}' - 1} 
   + \delta_{S_{\beta,2}, S_{\beta,2}' + 1}\right)
  \delta_{S_{\beta,4}, S_{\beta,4}'} \\
  & \qquad + \left(\delta_{S_{\beta,4}, S_{\beta,4}' - 1} 
   + \delta_{S_{\beta,4}, S_{\beta,4}' + 1}\right)
  \delta_{S_{\beta,2}, S_{\beta,2}'} \bigg] 
  \left[1 - \tanh
   \frac{\Delta\mathcal{F}_{\text{hop}}(\bm{S}_{\beta}';\bm{S}_{\beta})}{2
   k_{B} T} \right]  
 \end{split}
\end{equation}
\begin{equation}
 \frac{\Delta\mathcal{F}_{\text{hop}}(\bm{S}_{\beta}';\bm{S}_{\beta})}{k_{B}
  T} \equiv \frac{3}{2 N_{s} b^{2}}
  \left[ (\bm{R}_{S_{\beta,1}',S_{\beta,2}'} -
  \bm{R}_{S_{\beta,3}',S_{\beta,4}'})^{2}
  - (\bm{R}_{S_{\beta,1},S_{\beta,2}} -
  \bm{R}_{S_{\beta,3},S_{\beta,4}})^{2} \right]
\end{equation}
The master equation for the hopping can be written as
\begin{equation}
 \label{master_equation_exchange}
 \begin{split}
  \frac{\partial P(\lbrace \bm{R}_{i,k} \rbrace,\lbrace \bm{S}_{\alpha}
  \rbrace,Z;t)}{\partial t}
  & = \sum_{\beta = 1}^{Z} \sum_{\bm{S}_{\beta}'}
  \big[ W(\bm{S}_{\beta}|\bm{S}_{\beta}') P(\lbrace \bm{S}_{\alpha}'
  \rbrace)
  -  W(\bm{S}_{\beta}'|\bm{S}_{\beta}) P(\lbrace \bm{S}_{\alpha}\rbrace) \big]
  \\
  & \equiv \mathcal{L}_{\text{hop}} P
 \end{split} 
\end{equation}
Here we have introduced the time evolution operator for the
hopping process $\mathcal{L}_{\text{hop}}$.

Full dynamics of the system 
 is then described by the Langevin equation given by eq
 \eqref{langevin_equation_bead}, the reconstruction process
 (with the reconstruction rates \eqref{destruction_rate} and \eqref{construction_rate}), and the
 hopping process (with the hopping rate \eqref{hopping_rate}). All of these processes satisfy
 the detailed balance condition, and thus it is clear that the
 equilibrium state is realized as characterized by the probability
 distribution \eqref{equilibrium_probability_slip_spring_position_final_modified}.
The master equation of the system is expressed by combining the
Fokker-Planck and master equations, \eqref{fokker_planck_equation_bead},
\eqref{master_equation_reconstruction}, and \eqref{master_equation_exchange}.
\begin{equation}
 \label{master_equation}
  \frac{\partial P(\lbrace \bm{R}_{i,k} \rbrace,\lbrace \bm{S}_{\alpha}
  \rbrace,Z;t)}{\partial t}
  = \left[ \mathcal{L}_{\text{FP}} + \mathcal{L}_{\text{rc}}
    + \mathcal{L}_{\text{hop}} \right] P(\lbrace \bm{R}_{i,k} \rbrace,\lbrace \bm{S}_{\alpha}
  \rbrace,Z;t)
\end{equation}


\subsection{Relaxation modulus}
\label{Sec:Gt}
In this section, we will derive an explicit expression
 of the relaxation modulus tensor from the equilibrium distribution
 (eq \eqref{equilibrium_probability_slip_spring_position_final_modified})
 and the master equation (eq \eqref{master_equation}) via the linear response theory\cite{Risken}.

We consider the system is subjected to the weak external deformation field,
which is characterized by the time-dependent velocity gradient tensor
$\bm{\kappa}(t)$. Such a deformation gives an additional term to the
master equation \eqref{master_equation}, which will be treated as a
perturbation in the followings.
By adding the perturbation term, the master equation \eqref{master_equation} is modified as
\begin{equation}
 \label{master_equation_modified}
  \frac{\partial P(\lbrace \bm{R}_{i,k} \rbrace,\lbrace \bm{S}_{\alpha}
  \rbrace,Z;t)}{\partial t}
  = \left[ \mathcal{L}_{0} + \mathcal{L}_{1}(t) \right] P(\lbrace \bm{R}_{i,k} \rbrace,\lbrace \bm{S}_{\alpha}
  \rbrace,Z;t) 
\end{equation}
where $\mathcal{L}_{0}$ and $\mathcal{L}_{1}(t)$ are the equilibrium and
perturbation time evolution operators.
\begin{align}
 & \mathcal{L}_{0} \equiv \mathcal{L}_{\text{FP}} + \mathcal{L}_{\text{rc}}
    + \mathcal{L}_{\text{hop}} \\
 & \mathcal{L}_{1}(t) P \equiv - \sum_{i,k} \frac{\partial}{\partial
 \bm{R}_{i,k}} \cdot \left[ \bm{\kappa}(t) \cdot \bm{R}_{i,k} P \right]
\end{align}
As we mentioned, our dynamics model satisfies the detailed balance
condition and thus the following relation holds for $\mathcal{L}_{0}$
and the equilibrium distribution \eqref{equilibrium_probability_slip_spring_position_final_modified}.
\begin{equation}
 \mathcal{L}_{0} P_{\text{eq}}(\lbrace \bm{R}_{i,k} \rbrace,\lbrace \bm{S}_{\alpha}
  \rbrace,Z) = 0
\end{equation}

Up to the first order in the perturbation, eq
\eqref{master_equation_modified} can be formally integrated as
\begin{equation}
 \label{master_equation_modified_integrated}
  P(\lbrace \bm{R}_{i,k} \rbrace,\lbrace \bm{S}_{\alpha}
  \rbrace,Z;t)
  = P_{\text{eq}}(\lbrace \bm{R}_{i,k} \rbrace,\lbrace \bm{S}_{\alpha}
  \rbrace,Z) 
  + \int_{-\infty}^{t} dt' \, e^{(t - t') \mathcal{L}_{0}}\left[ \mathcal{L}_{1}(t') P_{\text{eq}}(\lbrace \bm{R}_{i,k} \rbrace,\lbrace \bm{S}_{\alpha}
  \rbrace,Z) \right]
\end{equation}
Then the ensemble average of the stress tensor at time $t$, $\bm{\sigma}(t)$, can be written as
\begin{equation}
 \label{linear_response_of_stress_tensor}
 \begin{split}
  \bm{\sigma}(t)
  & = \int d\lbrace \bm{R}_{i,k} \rbrace \sum_{Z = 0}^{\infty}
  \sum_{\lbrace \bm{S}_{\alpha} \rbrace} \hat{\bm{\sigma}} P(\lbrace \bm{R}_{i,k} \rbrace,\lbrace \bm{S}_{\alpha}
  \rbrace,Z;t) \\
  & = \langle \hat{\bm{\sigma}} \rangle_{\text{eq}} + \frac{1}{k_{B} T} \int_{-\infty}^{t} dt' \, \bigg\langle \hat{\bm{\sigma}}(t - t') \sum_{i,k} 
  \bigg[ 
  \frac{\partial \mathcal{F}(\lbrace \bm{R}_{i,k} \rbrace,\lbrace \bm{S}_{\alpha}
  \rbrace,Z)}{\partial \bm{R}_{i,k}}  \bm{R}_{i,k} - k_{B} T \bm{1} \bigg] \bigg\rangle_{\text{eq}} :
  \bm{\kappa}(t') \\
  & = \langle \hat{\bm{\sigma}} \rangle_{\text{eq}} + \frac{V}{k_{B} T} \int_{-\infty}^{t} dt' \, \big\langle \hat{\bm{\sigma}}(t - t')
  [ \hat{\bm{\sigma}}  + \hat{\bm{\sigma}}^{(v)}] \big\rangle_{\text{eq}} :
  \bm{\kappa}(t')
 \end{split}
\end{equation}
Here we have defined the time shifted operator as $\hat{\bm{\sigma}}(t) \equiv e^{t \mathcal{L}_{0}^{\dagger}}
\hat{\bm{\sigma}}$ ($\mathcal{L}_{0}^{\dagger}$ is the adjoint operator for
 $\mathcal{L}_{0}$). This represents the stress tensor at time
 $t$ after the reference time. Also, we have defined the virtual stress operator
$\hat{\bm{\sigma}}^{(v)}$ as
\begin{equation}
 \label{virtual_stress_tensor_definition}
 \begin{split}
  \hat{\bm{\sigma}}^{(v)}
  & \equiv \frac{1}{V} \bigg[ \sum_{\alpha = 1}^{Z} \frac{3 k_{B} T}{N_{s} b^{2}}
  (\bm{R}_{S_{\alpha,1},S_{\alpha,2}} - \bm{R}_{S_{\alpha,3},S_{\alpha,4}})
  (\bm{R}_{S_{\alpha,1},S_{\alpha,2}} - \bm{R}_{S_{\alpha,3},S_{\alpha,4}}) \\
  & \qquad - e^{\nu / k_{B} T}
  \sum_{i,k,j,l} \frac{3 k_{B} T}{N_{s} b^{2}}  (\bm{R}_{i,k} -
  \bm{R}_{j,l}) (\bm{R}_{i,k} - \bm{R}_{j,l}) \exp
\left[ - \frac{3 (\bm{R}_{i,k} - \bm{R}_{j,l})^{2}}{2 N_{s} b^{2}}
  \right]  \bigg]
 \end{split}
\end{equation}
The virtual stress represents the stress generated by slip-springs and
the repulsive interaction between beads.
The relaxation modulus tensor $\bm{G}(t)$ (which is a fourth order
tensor) can be defined for a small
deformation as follows.
\begin{equation}
 \label{relaxation_modulus_definition}
 \bm{\sigma}(t) -
   \langle \hat{\bm{\sigma}} \rangle_{\text{eq}}
   \equiv \int_{-\infty}^{t} dt' \, \bm{G}(t - t') : \bm{\kappa}(t')
\end{equation}
By comparing eqs \eqref{linear_response_of_stress_tensor} and
\eqref{relaxation_modulus_definition}, we obtain
\begin{equation}
\label{relaxation_modulus_expression}
 \bm{G}(t)
  = \frac{V}{k_{B} T} \big\langle \hat{\bm{\sigma}}(t)
  \big[ \hat{\bm{\sigma}}  + \hat{\bm{\sigma}}^{(v)} \big] \big\rangle_{\text{eq}}
\end{equation}
Eq \eqref{relaxation_modulus_expression} is similar to the linear
response formula obtained for single chain models\cite{Ramirez2007,Uneyama2011}
 where the necessity of the virtual stress tensor is
already known. 
However, the explicit form of the virtual stress tensor
\eqref{virtual_stress_tensor_definition} differs from one for single
chain models. In our model, the virtual
stress tensor has the contribution from the repulsive interaction
between beads.
Physically this is natural because the
repulsive interaction originates as the compensation of the attractive interaction
by the slip-springs.

{In eq \eqref{relaxation_modulus_expression}, we assume that only the stress
tensor $\hat{\bm{\sigma}}$ represents the stress tensor of the
system (eq {\eqref{stress_tensor_definition}}). As we mentioned, this assumption is based on the stress-optical
law, which is empirically known to hold for various polymeric materials. However, from
the view point of the virtual work method, it is also possible to employ
$\hat{\bm{\sigma}} + \hat{\bm{\sigma}}^{(v)}$ (which is conjugate to the
deformation) as the stress tensor of the system. If we employ the latter
expression, the ensemble average in eq
{\eqref{relaxation_modulus_expression}} is replaced by
$\langle [\hat{\bm{\sigma}}(t) + \hat{\bm{\sigma}}^{(v)}(t) ]
[ \hat{\bm{\sigma}}  + \hat{\bm{\sigma}}^{(v)} ]
\rangle_{\text{eq}}$. Then we have the following formula.}
\begin{equation}
\label{relaxation_modulus_expression_alternative}
 \bm{G}(t)
  = \frac{V}{k_{B} T} \big\langle \big[ \hat{\bm{\sigma}}(t)
   + \hat{\bm{\sigma}}^{(v)}(t) \big]
  \big[ \hat{\bm{\sigma}} + \hat{\bm{\sigma}}^{(v)} \big] \big\rangle_{\text{eq}}
\end{equation}
{(In the single chain slip-spring model, both of
these two different expressions give qualitatively similar relaxation moduli.}{\cite{Uneyama2011}} 
{ We expect that the situation is similar in our multi chain model.)
We will compare simulation results for eqs
{\eqref{relaxation_modulus_expression_alternative}}
and {\eqref{relaxation_modulus_expression_alternative}}, later.}

\subsection{Numerical scheme}

In this section, we show a numerical scheme for simulations
based on our multi-chain slip-spring model.
We choose the bead size $b$, $k_{B} T$, and $\zeta$ as the unit of
length, energy, and friction (so that the unit of time
 is $\tau_0 = \zeta b^2 / k_{B}T$). In the followings we set $b = 1$,
 $k_{B} T = 1$, and $\zeta = 1$.
By using the operator-splitting method, the formal solution of eq \eqref{master_equation} 
 can be approximated as
\begin{equation}
\label{master_equation_approximated}
 P(\lbrace \bm{R}_{i,k} \rbrace,\lbrace \bm{S}_{\alpha}
  \rbrace,Z;t + \Delta t)
  \approx e^{\Delta t \mathcal{L}_{\text{rc}}}
    e^{\Delta t \mathcal{L}_{\text{hop}}}
    e^{\Delta t \mathcal{L}_{\text{FP}}} P(\lbrace \bm{R}_{i,k} \rbrace,\lbrace \bm{S}_{\alpha}
  \rbrace,Z;t)
\end{equation}
Eq \eqref{master_equation_approximated} corresponds to a numerical scheme
with three substeps and the integration time step $\Delta t$.
That is, the time evolution from time $t$ to time $t + \Delta t$ is
simulated by performing the Langevin dynamics
of the beads, the hopping dynamics of the slip-springs, and the reconstruction
of the slip-springs. (These steps are iterated sequentially.)

For the integration of the Langevin equation, we employ the explicit
Euler scheme. The Langevin equation for the chain dynamics 
can be discretized as
\begin{equation}
 \bm{R}_{i,k}(t + \Delta t) \approx - \Delta t
  \frac{\partial \mathcal{F}(\lbrace \bm{R}_{i,k} \rbrace,\lbrace \bm{S}_{\alpha}
  \rbrace,Z)}{\partial \bm{R}_{i,k}} + \sqrt{2 \Delta t} \bm{w}_{i,k}
\end{equation}
Here $\bm{w}_{i,k}$ is the Gaussian random number vector.

The hopping dynamics of slip-spring is described by the change of 
slip-spring indices as mentioned above.
When $\Delta t$ is sufficiently small, the index of the $\alpha$-th
slip-spring, $S_{\alpha,\lambda}$ ($\lambda = 2,4$) is changed
as $S_{\alpha,\lambda} \to S_{\alpha,\lambda} \pm 1$ by the following
cumulative probability.
\begin{equation}
 \Psi_{\lambda\pm} = \frac{\Delta t}{\zeta_{s}} \left[1 - \tanh
   \frac{\Delta\mathcal{F}_{\lambda\pm}}{2} \right] \qquad (\lambda = 2,4)
\end{equation}
Here $\Delta\mathcal{F}_{\lambda\pm}$ is the free energy difference given by
\begin{align}
 & \Delta\mathcal{F}_{2\pm}
  \equiv \frac{3}{2 N_{s}}
  \left[ (\bm{R}_{S_{\alpha,1},S_{\alpha,2} \pm 1} -
  \bm{R}_{S_{\alpha,3},S_{\alpha,4}})^{2}
  - (\bm{R}_{S_{\alpha,1},S_{\alpha,2}} -
  \bm{R}_{S_{\alpha,3},S_{\alpha,4}})^{2} \right] \\
 & \Delta\mathcal{F}_{4\pm}
  \equiv \frac{3}{2 N_{s}}
  \left[ (\bm{R}_{S_{\alpha,1},S_{\alpha,2}} -
  \bm{R}_{S_{\alpha,3},S_{\alpha,4} \pm 1})^{2}
  - (\bm{R}_{S_{\alpha,1},S_{\alpha,2}} -
  \bm{R}_{S_{\alpha,3},S_{\alpha,4}})^{2} \right]
\end{align}

The reconstruction of the slip-springs is performed as follows.
When an end of the $\alpha$-th slip-spring is at a chain end, the slip-spring
is {destroyed} with the following cumulative probability.
\begin{equation}
 \Psi_{-}
  = \frac{\Delta t}{\zeta_{s}}
\end{equation}
When the slip-spring is {destroyed}, the index $\alpha$ for the other
 slip-springs is rearranged to realize $\alpha= 1,2,3,\dots,Z$ without
 vacant number. (This rearrangement is expressed by the exchange
 operator in eq \eqref{destruction_rate}.)
After the attempts of destruction for all slip-springs, construction of 
 a new slip-spring is attempted. This construction step is made by a Monte Carlo
 sampling scheme. A new slip-spring is virtually generated and its end is
 attached to a chain end. Another end is attached to one of the
 surrounding beads which is chosen randomly. Thus a new slip-spring
 index $\bm{S}$ is generated.
This virtual slip-spring is accepted as a newly generated slip-spring with the following probability.
\begin{equation}
 \Psi_{+} = 
  \exp \left[ - \frac{3 (\bm{R}_{S_{1},S_{2}} - \bm{R}_{S_{3},S_{4}})^{2}}{2 N_{s}} \right]
\end{equation}
If accepted, we set $\bm{S}_{Z + 1} = \bm{S}$ and increase $Z \to Z + 1$.
This Monte Carlo sampling is made for 
\begin{equation}
 \bar{K} \equiv 4 M^{2} N \frac{\Delta t}{\zeta_{s}} e^{\nu}
\end{equation}
times on average, where the factor $4 M^{2} N$ is the total number of 
 possible connections factorized by the number of ends for a slip-spring, 2, the
  total number of chain ends, $2 M$, and the total number of beads, $MN$. 
As a simple scheme, we assume that $K$ only takes the floor or ceiling of $\bar{K}$,
($\lfloor \bar{K} \rfloor$ or $\lceil \bar{K} \rceil$). The
 probability that we have $K$ trials at a certain construction step is
 given by
\begin{align}
 & P(K = \lfloor \bar{K} \rfloor)
 = \lceil \bar{K} \rceil - \bar{K} \\
 & P(K = \lceil \bar{K} \rceil )
 = \bar{K}
 - \lfloor \bar{K} \rfloor
 = 1 - P(K = \lfloor \bar{K} \rfloor)
\end{align}
This guarantees that the average sampling number becomes $\langle K
\rangle = \bar{K}$, and the numbers of the slip-spring construction and destruction
balance in equilibrium.

The number of Monte Carlo sampling for construction can be
significantly reduced if we exclude the constructions for very low
acceptance probabilities. The acceptance probability decreases
considerably if the stretch of a slip-spring becomes large. Thus, we 
limit the newly constructed slip-springs to be sampled only inside a
certain cut-off size.
We introduce a cut-off $r_{c} \equiv \sqrt{C_{0}^{2} N_{s} / 3}$ to restrict
 the sampling for construction of slip-spring with its length {shorter} than $r_{c}$. 
The cut-off parameter $C_{0}$ is determined to obey $e^{-C_{0}^{2} / 2} \ll 1$
 and practically this condition is satisfied if $C_{0} \gtrsim 3$.
The average number of trials for every $\Delta t$ is written as
\begin{equation}
 \label{average_number_of_trials_with_cutoff}
 \bar{K}
 = 4 M \tilde{N}_{0}(C_{0})
  \frac{\Delta t}{\zeta_{s}} e^{\nu}
\end{equation}
Here $\tilde{N}_{0}(C_{0})$ is the average number of beads located inside the 
 distance $\sqrt{C_{0}^{2} N_{s} / 3}$ from the subjected chain end.
$\tilde{N}_{0}(C_{0})$ is written as
\begin{equation}
 \label{average_number_of_active_beads_cutoff}
 \begin{split}
 \tilde{N}_{0}(C_{0})
  & \approx \int_{|\bm{r}| < r_{c}} d\bm{r} \, \bigg\langle 
  \sum_{k = 1}^{N} \delta(\bm{R}_{i,1} - \bm{R}_{i,k} - \bm{r})  \bigg\rangle_{\text{eq}} + \frac{4 \pi r_{c}^{3}}{3} \frac{(M - 1) N}{V} \\
  & = 1 + \sum_{k = 1}^{N - 1}
  \bigg[
  \erf \sqrt{\frac{C_{0}^{2} N_{s}}{2 k}}
  - \sqrt{\frac{2 C_{0}^{2} N_{s}}{\pi k}}
  \exp\bigg(- \frac{C_{0}^{2} N_{s}}{2 k} \bigg) \bigg] \\
  & \qquad + \frac{4 \pi}{3} \left(\frac{C_{0}^{2}
  N_{s}}{3}\right)^{3/2} \bigg(\rho_{0} - \frac{N}{V} \bigg)
 \end{split}
\end{equation}
Eq \eqref{average_number_of_active_beads_cutoff} can be numerically
evaluated if the parameters (such as $C_{0}$ and $N_{s}$) are given, and as long as the parameters
are unchanged during the simulation, $\bar{K}$ can be treated as constant. 
With this cut-off, the number of trials (for constant $\rho_{0}$) is 
$O(\bar{K}) = O(M)$ which is much smaller than that without cut-off, $O(\bar{K}) = O(M^{2})$.

\subsection{Comparison with earlier models}
\label{Sec:Comp}
In this section, to clarify the position of the proposed model,
we compare our model with couples of similar models for entangled polymers.

It has been rather established that Kremer-Grest type coarse-grained molecular dynamics simulation\cite{Kremer} is the standard way 
to reproduce polymer dynamics including entangled systems. 
In this approach, the multi-chain dynamics is solved with the excluded volume interaction that guarantees uncrossability between chains. 
On the other hand, in our model the interaction between chains is very
soft and the entanglement effect is introduced a priori by the slip-springs.
 Since the relation between the contacts among chains in Kremer-Grest simulations and the entanglement used 
 in the entanglement-based models (such as our model) has not been clarified, 
 our slip-spring reconstruction rules can not be related to the dynamics in Kremer-Grest simulations at this time being. 
For instance, it has been reported that in Kremer-Grest simulations the long-lived contacts between two chains are constructed
 not only around the chain ends but also interior of the chain\cite{GaoWeiner1995,YamamotoOnuki}.
But in the presented model (and the other entanglement-based models) assumes that the entanglement reconstruction occurs
only at the chain ends. It should be remarked, however,
the resultant chain dynamics is similar to each other as shown later. 

Padding and Briels\cite{Twentangle} have proposed a smart approach
 (referred as the TWENTANGLEMENT model) to deal with the uncrossability among chains
 without the excluded volume interaction. 
In the TWENTANGLEMENT model, a crossing event between bonds
 (which polymer chains consists of) is mathematically detected, and if the crossing occurs
 the force between segments is generated, to avoid chains cross freely.
Due to the absence of the excluded volume interaction, their model is much coarse-grained than Kremer-Grest model
 and rather close to our model. 
On the other hand, the basic idea on the entanglement is the common for
the Kremer-Grest and TWENTANGLEMENT models. Namely, both models do not
require any artificial objects
which represent the uncrossability (such as slip-springs in our model). 
We should also mention that there are no artificial attraction between
chains in the TWENTANGLEMENT model, and thus the repulsive interaction is not required
unlike our model.
In these aspects, the TWENTANGLEMENT model is located in between the
Kremer-Grest model and our model.

Masubuchi et al\cite{JCP2001} have developed another multi-chain model
called the primitive chain network (PCN) model as mentioned in the introduction. 
The model presented in this work and the PCN model are similar to each other; in both models, phantom chains are connected to form a network, 
 and the entanglement is mimicked by bundling of segments rather artificially.
The reconstruction of entanglement is assumed to occur only at the chain ends as considered in the tube theory. 
One fundamental difference is the level of description. 
In the PCN model, only the number of Kuhn segments between entanglements
is used and the position of each segment is not monitored. 
Thus, the PCN model cannot deal with the dynamics and structure in the time and length scales below those of entanglement. 
On the contrary, in the present model the dynamics of segments between
entanglements is considered explicitly. 
This difference gives a difference in computational costs. The
computational cost of the PCN is much smaller than one of the present model, as shown later.
Another difference is the thermodynamic consistency, which is fully
considered in the present model while not in the PCN model. 
The reconstruction of entanglements in the PCN model does not fulfill the detailed balance condition, for example\cite{JCP2010}. 
Finally, we point that the strength of dynamical constraint is different; 
the slip-link employed in PCN corresponds to the limit of $N_s=0$ for the slip-spring of the present model. 
This difference may affect some properties such as the orientation tensor under deformations\cite{Furuichi2008}.

As mentioned in the introduction, the presented model is the many chain
version of the Likhtman's single chain model\cite{Likhtman2005} where 
several slip-springs are connected to a single Rouse chain.
In the Likhtman's model, one end of a slip-spring can slide along the
chain contour while another end is fixed in space.
It the present model, on the other hand, both ends of a slip-spring can
slide along chains.
As we shall discuss later, this difference affects the effective strength of the dynamical constraint to the chain dynamics. 
Another difference between the two models is that the number of springs (standing for entanglements) is assumed to be constant
 in Likhtman's model while it fluctuates in the present model since it is controlled via the chemical potential. 
It is also pointed that there are a couple of differences in the sliding rule for slip-springs along the chain. 
As we mentioned, our slip-spring hops between segments while Likhtman's
slip-spring actually slides on the bond between segments.
(However, the difference between the hopping on beads and sliding along the bond
between segments seems to have minor effects to dynamical properties, as
judged from single chain simulation results\cite{Uneyama2011}.)
Furthermore, in Likhtman's model the slip-springs are allowed neither to change their order along the chain 
 nor to overlap with each other, while in our model the motion of slip-spring is completely independent. 

It should be noted that there have
been proposed several single chain slip-link models with the thermodynamic consistency.
Schieber et al\cite{Schieber2003,Nair,Renat2009} have proposed such models where the state variables are chosen as the number of entanglement, 
the position of entanglement and the number of monomers between entanglements. 
This choice of state variables is similar to one of the PCN model, but
fundamental difference is that these models have well-defined free
energy free energy of the system. The dynamic equations or transition rates
are derived from the free energy and the detailed balance condition.
The chemical potential controls the fluctuation of number of entanglements and this strategy is employed in our model. 
Due to the nature of the single chain model, it is intrinsically difficult to deal with the effect of surrounding chains. 
(Although several attempts have been made to overcome this
difficulty\cite{Ekaterina,Renat}, some additional and non-trivial assumptions are
required to mimic multi chain effects.)

\subsection{Simulations}
Monodisperse linear polymers were examined where bead number per chain $N$ was varied 
 {from 4 to 64}. The total number of beads in the system, $M N$, is
 fixed to be constant so that the bead number density was constant at {$\rho_0=4$}. 
Periodic boundary condition was utilized with the box dimension
 at {$8^3$}. 
The spring strength parameter $N_{s}$ for slip-springs was set as $N_s=0.5$. The number density of slip-springs
 was chosen as $\phi=0.5$.
{Conceptually, this slip-spring density give a certain plateau
 modulus but we have not yet obtained the relation between the plateau
 modulus and these parameters.} 
The cut-off parameter and the corresponding cut-off distance were $C_0^2=10$ and $r_c^2=1.29$, respectively.
The friction coefficient for the slip-spring was set as $\zeta_s=\zeta$.
{For the numerical calculation $\Delta t$ was chosen as $0.01$, after we checked reasonable numerical convergence for $\Delta t < 0.02$.}

\section{Results and Discussion}

\subsection{Statics}

\begin{figure}[p]
\newpage
\begin{center}\includegraphics[width=10cm]{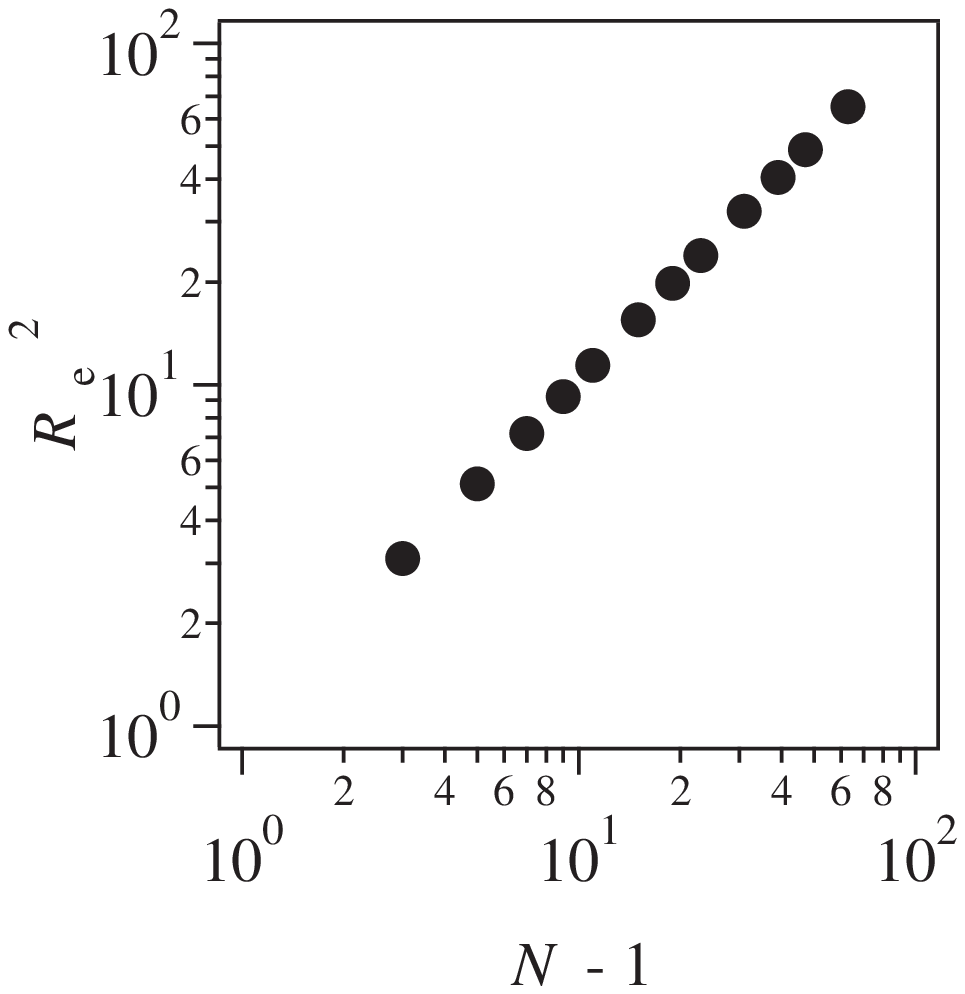}\end{center}
\caption
[ Squared end-to-end distance plotted against bond number.]
{T. Uneyama and Y. Masubuchi to {\it J. Chem. Phys}}
\label{Fig:Re2}
\end{figure}

\begin{figure}[p]
\newpage
\begin{center}\includegraphics[width=10cm]{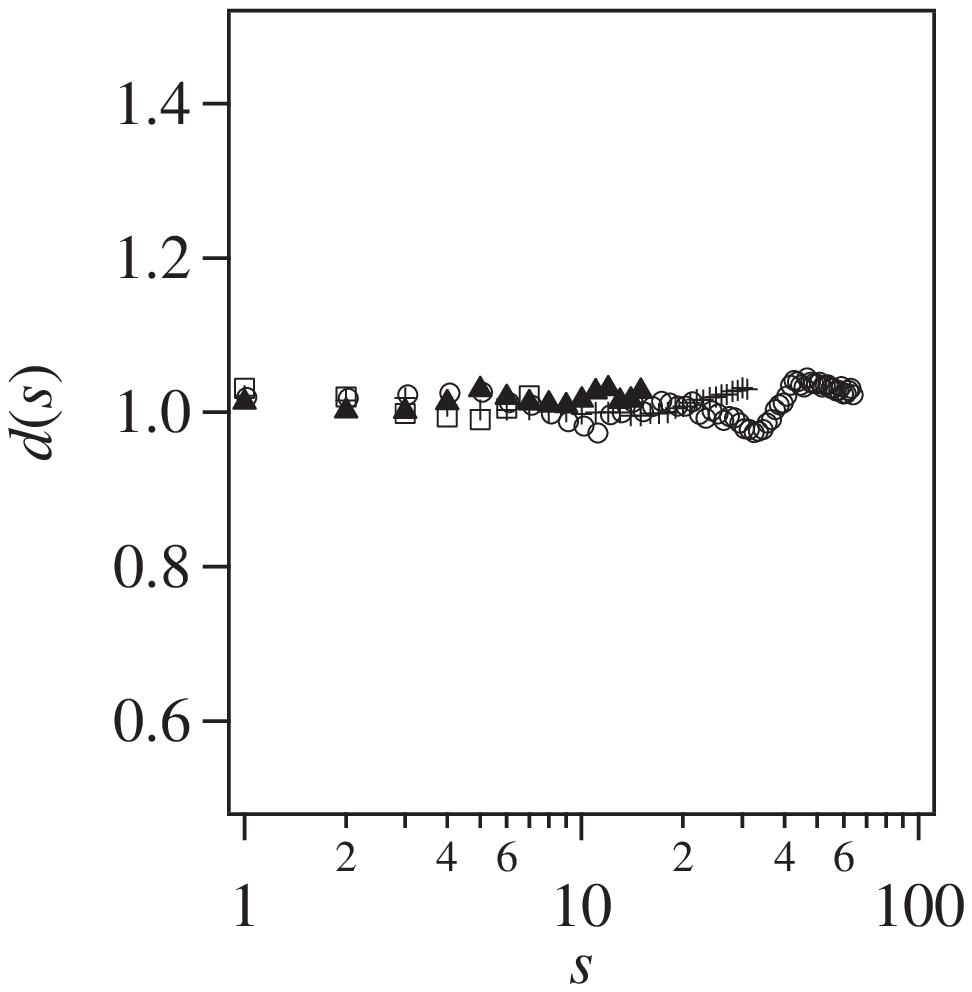}\end{center}
\caption
[ Internal distance factor $d(s)$ for $N=8, 16, 32$ and $64$ indicated by square, triangle, cross and circle, respectively.]
{T. Uneyama and Y. Masubuchi to {\it J. Chem. Phys}}
\label{Fig:ds}
\end{figure}

In this subsection static properties of the system is examined to show
 the consistency between simulation data and the equilibrium distribution function
 given by eq \eqref{equilibrium_probability_slip_spring_position_final_modified}. 
Figure \ref{Fig:Re2} shows the bond number $(N-1)$ dependence of squared end-to-end distance $R_e^2$ to report that 
 the scaling obeys the Gaussian chain statistics. 
%
%
The internal chain structure is examined via the internal distance factor $d(s)$ defined as
$d(s)=\langle ( \bm{R}_{i,k+s}-\bm{R}_{i,k})^{2} \rangle / s$, and shown in Figure \ref{Fig:ds}. 
$d(s)$ is reasonably close to unity independently of $s$ and $N$ as expected for  
 the Gaussian chain statistics. These results demonstrate that the attractive
 force induced by the slip-springs is correctly compensated by the repulsive interaction.
A similar attempt has been made for PCN model where the soft core
 repulsive interaction was introduced between slip-links\cite{Okuda},
 but a precise control of
 the chain dimension was difficult due to the lack of free energy expression
 for PCN model. (Indeed, it seems practically impossible to introduce
 the repulsive potential which exactly cancels the artificial attractive
 interaction in slip-link models such as the PCN model\cite{UneyamaJPSB2011}.)
%
%

Figure \ref{Fig:Zc} shows the distribution of slip-spring number per chain $Z_c$, $P(Z_c)$. 
The grand canonical type treatment of the slip-springs for single chain
models predict the Poisson distribution for the number of
slip-springs on a chain\cite{SchieberJCP2003,Uneyama2011}.
Indeed, the results shown by symbols reasonably close to the Poisson distribution drawn by solid curves where the average value
 of $Z_c$ is given as $2, 4, 8$ and $16$ for the examined chains with
 $N=8, 16, 32$ and $64$,
for the simulation parameter $N_e=4$ (calculated from 
 the slip-spring density and the bead density as $N_e=2\phi\rho_0$). 
%
%

\subsection{Dynamics}
\begin{figure}[p]
\newpage
\begin{center}\includegraphics[width=10cm]{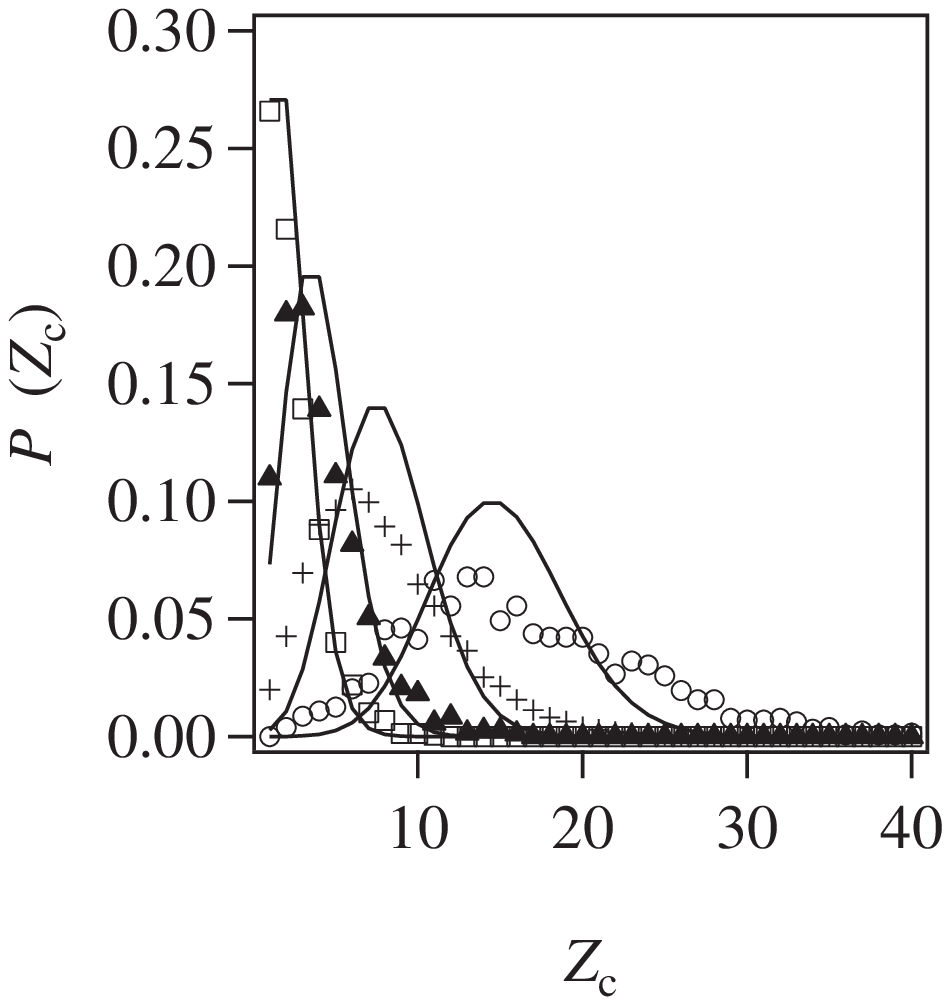}\end{center}
\caption
[ Distribution of slip-spring number per chain $Z_c$ for $N=8, 16, 32$ and $64$ indicated by square, triangle, cross and circle, respectively.
Poisson distributions with the average values of $2, 4, 8$ and $16$ are shown by solid curves.]
{T. Uneyama and Y. Masubuchi to {\it J. Chem. Phys}}
\label{Fig:Zc}
\end{figure}

\begin{figure}[p]
\newpage
\begin{center}\includegraphics[width=12cm]{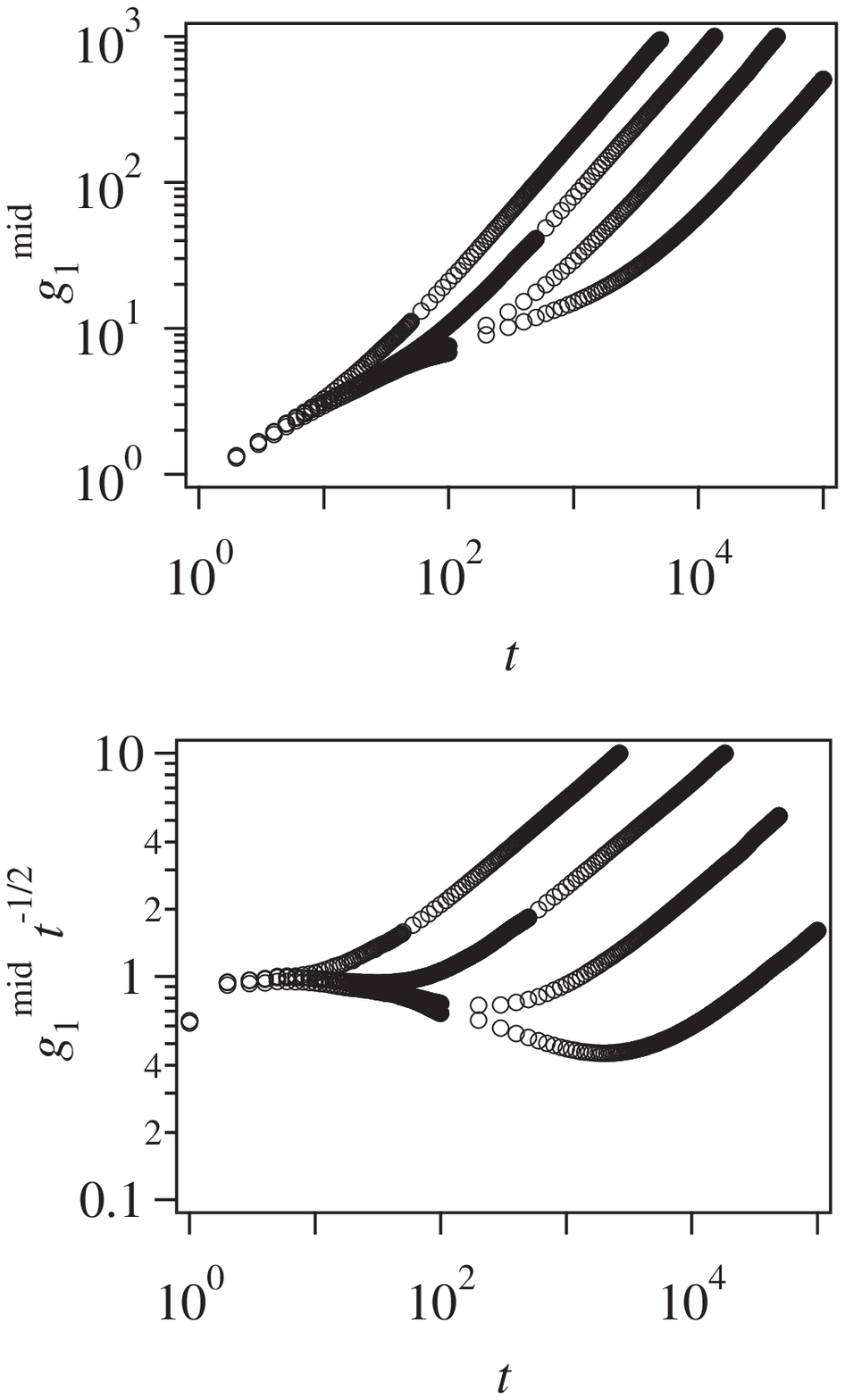}\end{center}
\caption
[ Mean square displacement of the central bead for $N=8, 16, 32$ and $64$ from left to right.]
{T. Uneyama and Y. Masubuchi to {\it J. Chem. Phys}}
\label{Fig:MSD}
\end{figure}

\begin{figure}[p]
\newpage
\begin{center}\includegraphics[width=15cm]{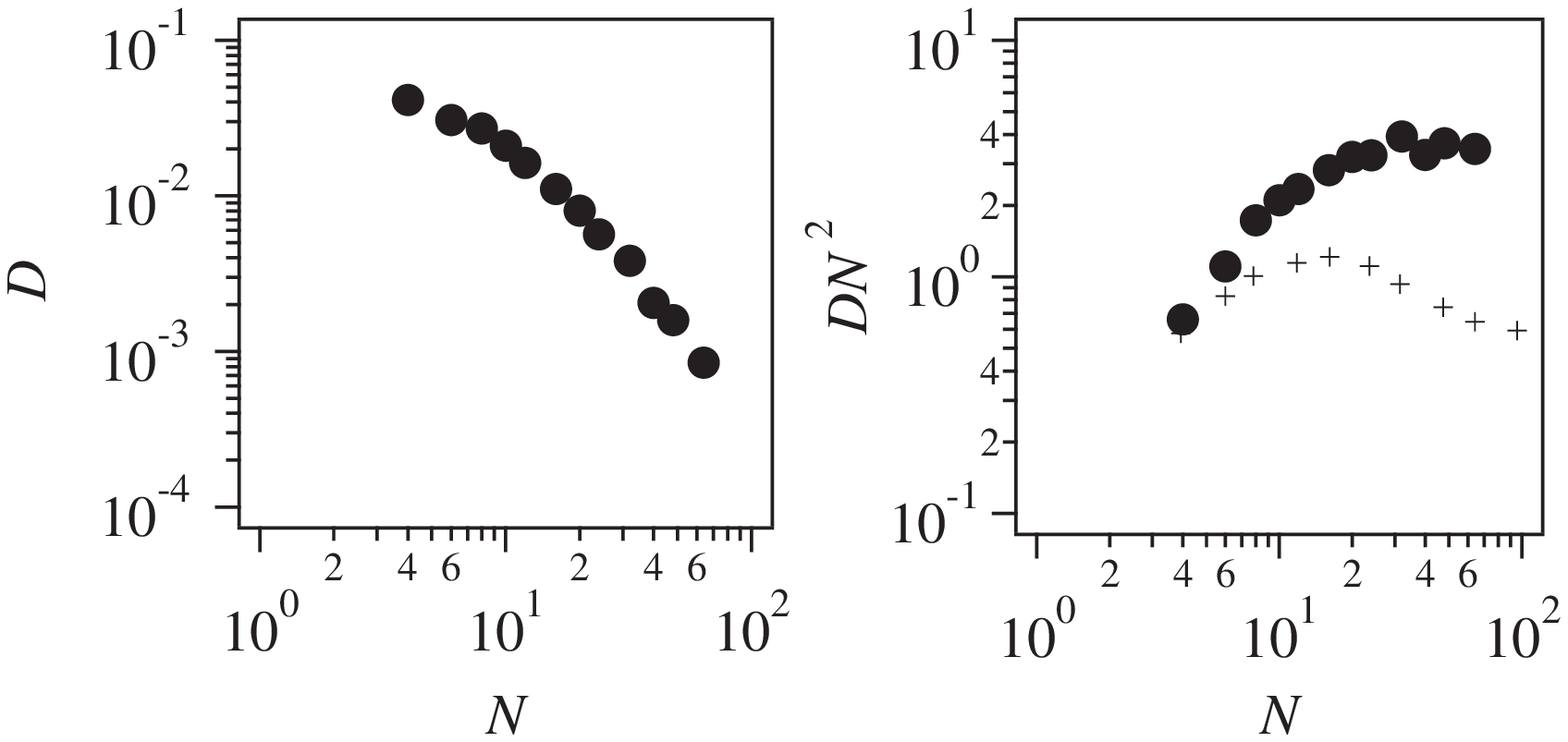}\end{center}
\caption
[ Diffusion coefficient $D$ plotted against bead number $N$. Right panel shows comparison with the single chain model\cite{Likhtman2005} indicated by cross.]
{T. Uneyama and Y. Masubuchi to {\it J. Chem. Phys}}
\label{Fig:D}
\end{figure}

\begin{figure}[p]
\newpage
\begin{center}\includegraphics[width=15cm]{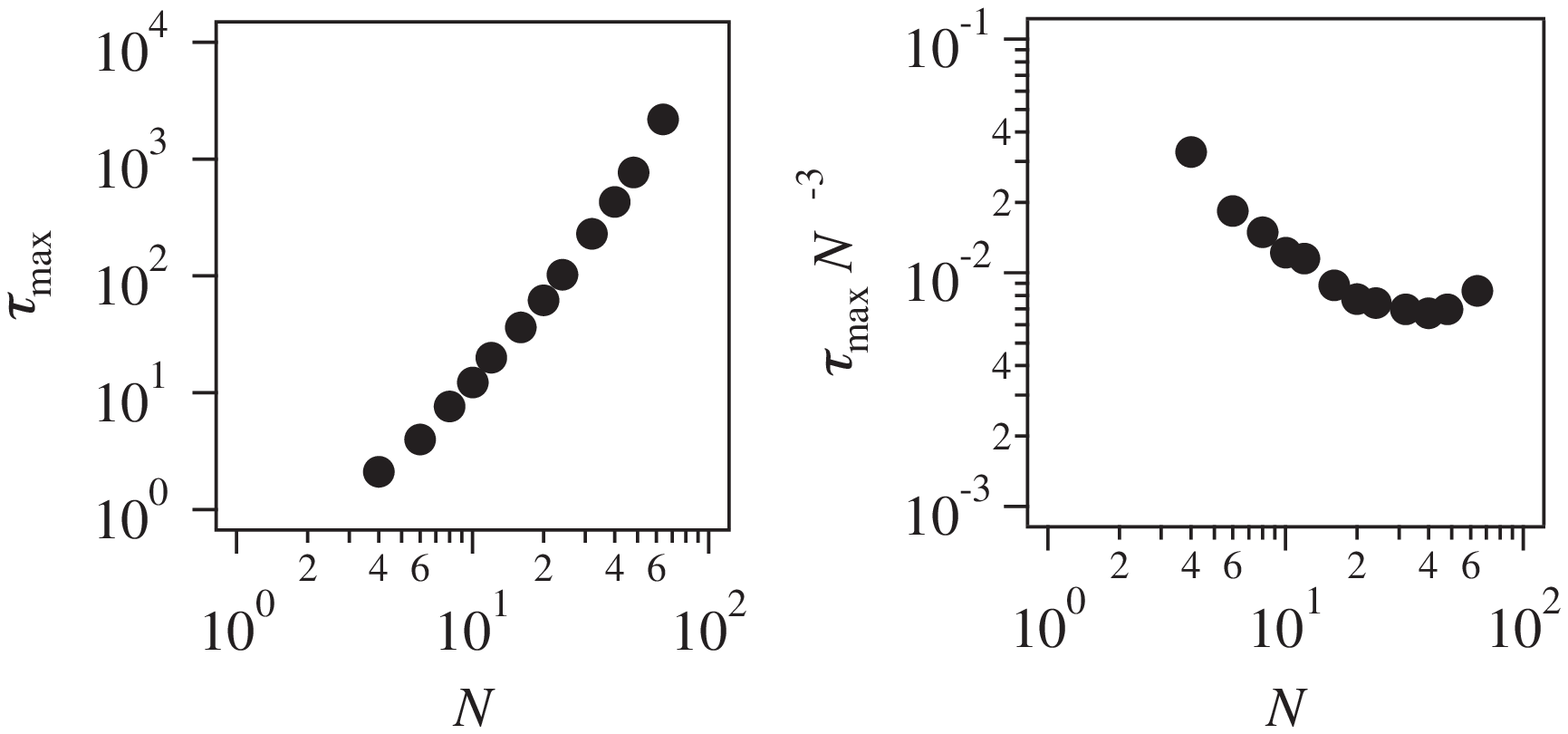}\end{center}
\caption
[ Relaxation time for end-to-end vector $\tau_{\text{max}}$ plotted against bead number $N$.]
{T. Uneyama and Y. Masubuchi to {\it J. Chem. Phys}}
\label{Fig:tau}
\end{figure}

Figure \ref{Fig:MSD} shows the mean square displacement of the central bead in the chain, $g_1^{\text{mid}}(t)$. 
To see the effect of the entanglement clearly, the data divided by
the Rouse behavior in internal time range
$g_1^{\text{mid}}(t) t^{-1/2}$ is also shown. From these plots, it is apparent that the entanglement effect appears after $t \sim 10$
 where the negative slope starts in $g_1^{\text{mid}}(t) t^{-1/2}$. In this respect, the short chain with $N=8$ does not show the
 entangled behavior, in spite of the fact that there exist two slip-springs per chain on average as shown in Fig \ref{Fig:Zc}.
%
%
Figure \ref{Fig:D} shows the diffusion coefficient $D$ against the bead
number $N$.
To see the scaling behavior clearly,
 $DN^2$ is also shown. 
The Rouse behavior $D\propto N^{-1}$ is observed below $N=8$, which is
consistent with what observed for $g_1^{\text{mid}}(t)$. 
For longer chains, the $N$ dependence is not that strong and actually the power law exponent does not largely deviate from $-2$
  even for $N=64$, which is somewhat weaker than experimental results for well entangled systems\cite{Lodge}. 
In comparison with the literature data (shown by cross in the right panel) for single chain slip-spring model\cite{Likhtman2005} with similar parameter setting
  (except the difference in $\zeta_s$ that is $1$ for our simulation while it is $0.1$ for the single chain data), it is suggested
  that the dynamical constraint in the present model is weaker than that in the single chain model. 
%
%
We expect that this is due to the difference in the hopping (sliding) kinetics of slip-springs.
In our model the both ends of slip-springs are mobile while in the
single chain models one end is anchored, as we mentioned in Section \ref{Sec:Comp}.
This difference will result in a weaker constraint to the chain in our model.
Indeed, similar behavior is observed in the $N$ dependence of the
longest relaxation time for the end-to-end vector, $\tau_{\text{max}}$, 
 shown in Fig \ref{Fig:tau}. Especially, $\tau_{\text{max}}N^{-3}$
 indicates that the large $N$ chains examined here are still
 in the transitional region between unentangled and entangled regimes. 
%
%

\begin{figure}[p]
\newpage
\begin{center}\includegraphics[width=10cm]{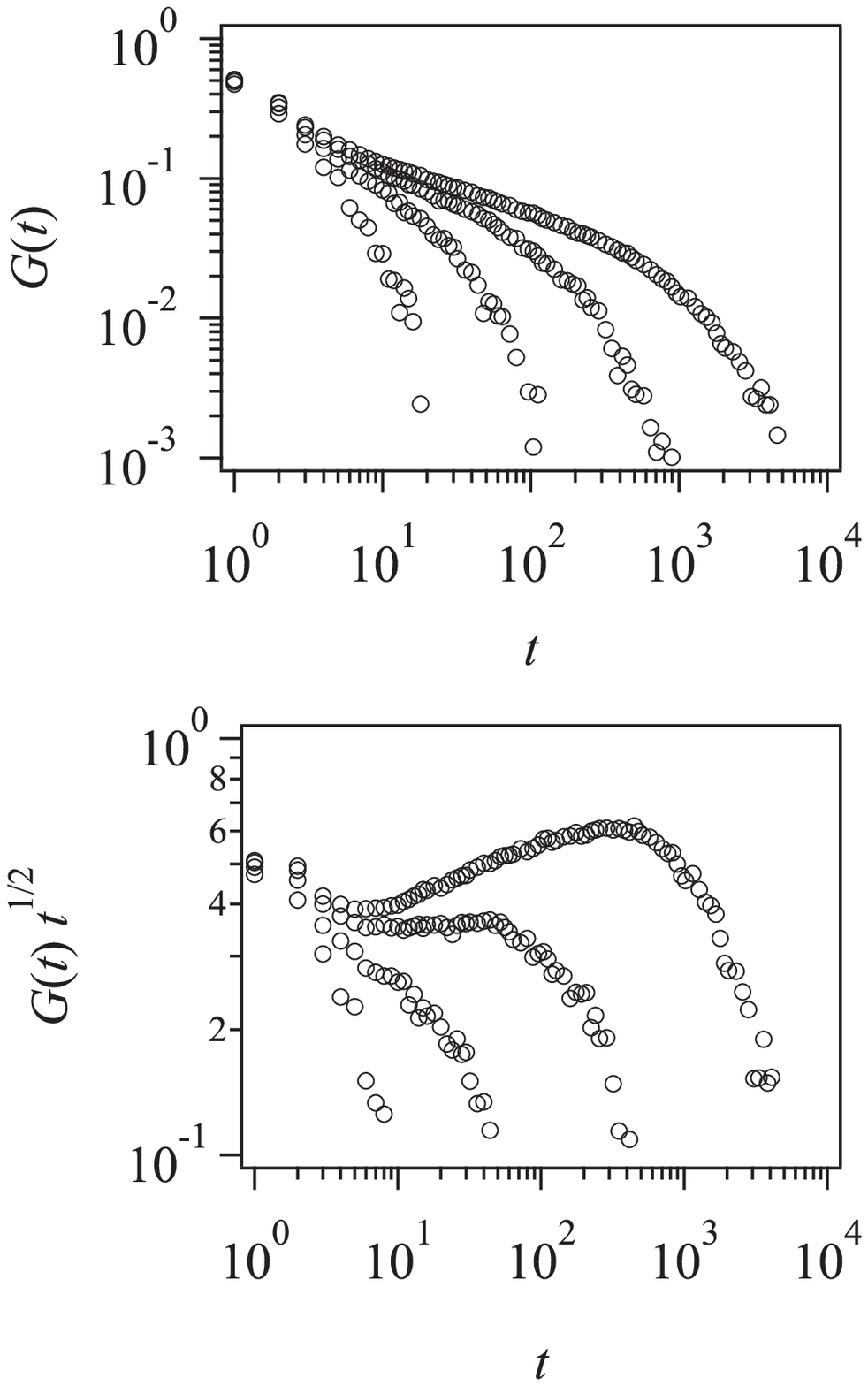}\end{center}
\caption
[ Linear relaxation modulus for $N=8, 16, 32$ and $64$ from left to right.]
{T. Uneyama and Y. Masubuchi to {\it J. Chem. Phys}}
\label{Fig:Gt}
\end{figure}

Figure \ref{Fig:Gt} shows the relaxation modulus $G(t)$ calculated by the linear response theory shown in Sec. \ref{Sec:Gt}.
Here, $G(t)$ is normalized by the unit modulus defined as $G_0\equiv \rho RT / M_0$\cite{Likhtman2007}. 
As in the case of the mean square displacement, we also show the data
divided by the Rouse behavior, $G(t)
t^{1/2}$.
As we have observed for $g_1^{\text{mid}}(t)$,
the entanglement effect (the increase of $G(t) t^{1/2}$ in time) is not observed in
short chains. 
On the other hand, for longer chains $G(t)$ apparently deviates from the Rouse relaxation to show the plateau, as expected.
We note that the unit modulus $G_0$ is not equal to the plateau modulus $G_N$ and actually $G_N$ is much smaller than $G_0$.
(This is because $G_{0}$ is a characteristic modulus for the bead scale,
and it reflects the short time relaxation modes.)
For the single chain model it is reported that $G_N \approx
0.1G_0$\cite{Likhtman2005}.
Our results for long chains are similar to this relation.
Of course, the relation between $G_{N}$ and $G_{0}$ depends on the model
and various parameters.
The direct comparison of our result with the single chain
simulation results seems to be difficult.
%
%

\begin{figure}[p]
\newpage
\begin{center}\includegraphics[width=10cm]{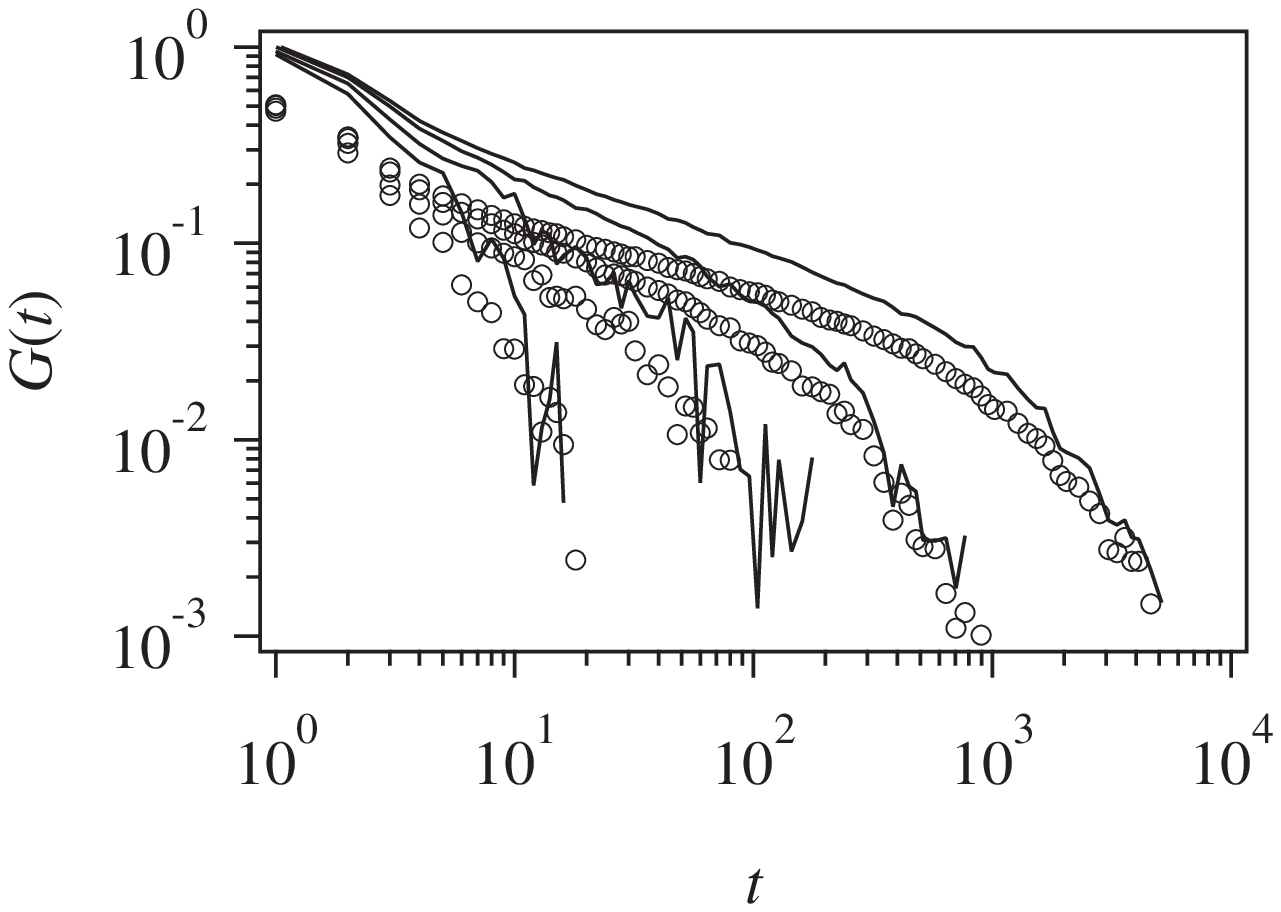}\end{center}
\caption
[ Linear relaxation moduli for $N = 8, 16, 32$ and 64, calculated with two different
 expressions of the stress tensor (eqs
{\eqref{relaxation_modulus_expression}} shown by symbol and
{\eqref{relaxation_modulus_expression_alternative}} shown by solid line).]
{T. Uneyama and Y. Masubuchi to {\it J. Chem. Phys}}
\label{Fig:Gt_virtual}
\end{figure}

As we mentioned, there are two possible expressions for the stress
tensor of the system. To investigate how
the expression of the stress tensor affects the rheological properties,
here we compare the shear relaxation moduli calculated with two
different expressions. Figure {\ref{Fig:Gt_virtual}} shows the shear
relaxation moduli for $N = 8, 16, 32$ and 64 calculated by eqs
{\eqref{relaxation_modulus_expression}} (symbol) and
{\eqref{relaxation_modulus_expression_alternative}} (solid curve). Although the shear
relaxation moduli data by eqs {\eqref{relaxation_modulus_expression}} and
{\eqref{relaxation_modulus_expression_alternative}} are quantitatively
different, they are qualitatively similar, as reported in the single chain model.\cite{Uneyama2011}

%
%

\begin{figure}[p]
\newpage
\begin{center}\includegraphics[width=10cm]{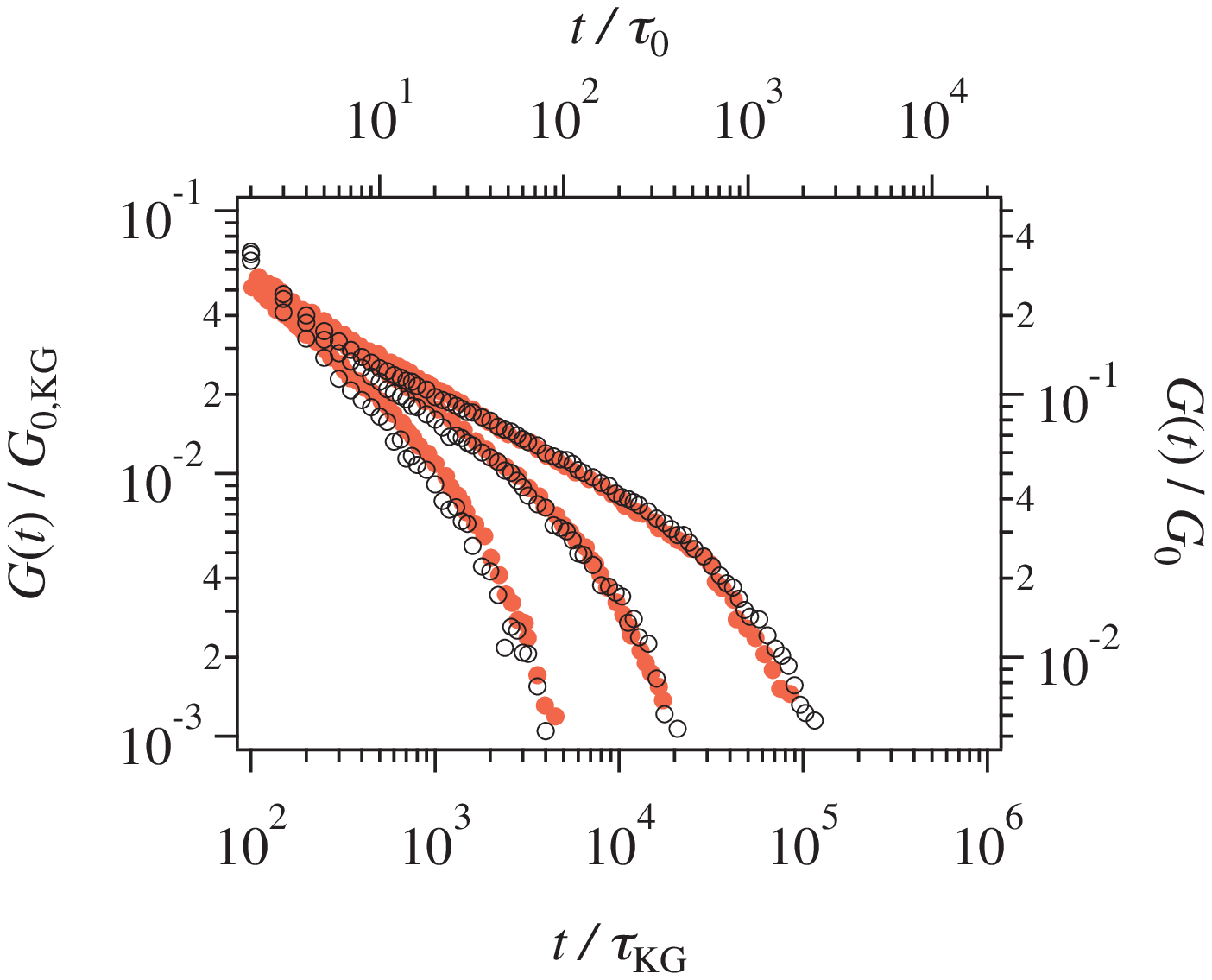}\end{center}
\caption
[ Linear relaxation modulus in comparison with the Kremer-Grest simulations. For the Kremer-Grest simulations the bead numbers are 50, 100 and 200 and the data taken from Ref. \cite{Likhtman2007} are shown by filled symbols according to left and bottom axes. For the present model the bead numbers are 16, 32 and 64 
 from left to right and the data are shown by unfilled symbols according to right and top axes. ]
{T. Uneyama and Y. Masubuchi to {\it J. Chem. Phys}}
\label{Fig:Gtcomp}
\end{figure}

Figure \ref{Fig:Gtcomp} shows a comparison with the Kremer-Grest simulations\cite{Likhtman2007} on $G(t)$. 
Here, $G(t)$ of our simulation is calculated by eq {\eqref{relaxation_modulus_expression}}, but
 similar fitting can be realized for $G(t)$ obtained by eq \eqref{relaxation_modulus_expression_alternative} as well,
 if the parameters are adequately tuned (not shown here).
There are small discrepancies in the short time region due to the difference in the level of coarse-graining,
 i.e., the number of Rouse modes.
This discrepancy may be eliminated if $N_e$ value is increased as reported in the single chain model.
For this fitting, we choose the parameters as follows.
\begin{equation}
 \label{relation_between_kg_and_slip_spring_parameters}
M_0=3.1, \quad \tau_0=50 \tau_{\text{KG}}, \quad G_0= 0.2 G_{0,\text{KG}}
\end{equation}
Here, $M_0$ is the number of beads (molecular weight) of the Kremer-Grest
model which corresponds to one
bead in our model, $\tau_{\text{KG}}$ is the unit time in the Kremer-Grest simulation (the
 standard time scale for particles with the Lennard-Jones potential), and $G_{0,\text{KG}}$ is the
 unit modulus for the Kremer-Grest model\cite{Likhtman2007}.
This fitting demonstrates that the dynamical constraint given by 
 the slip-spring is relatively weak, and this result is consistent with
 results obtained by $D$ and $\tau_{\text{max}}$. For instance, it has
 been reported that the entanglement number for the Kremer-Grest
  chain with 200 beads is around 5\cite{Likhtman2007}. On the other hand, the corresponding chain in our model is $N=64$ that has $16$ slip-springs
  per chain on average, as shown in Fig \ref{Fig:Zc}.
Since the $G(t)$ data compared here are in the transitional regime
and the contribution from the Rouse relaxation is
relatively large, the fitting results (eq
\eqref{relation_between_kg_and_slip_spring_parameters}) may be different
for well-entangled systems.
Nevertheless, the fact that $N_e$ is smaller than the effective
entanglement bead number is informative.
For this specific case, the effective entanglement bead number in our
model is roughly estimated as $3N_e$. 
A similar result have been reported for the PCN model\cite{JCP2003}, and such a
discrepancy is expected to be the nature of randomly connected multi
chain network.
From the similarity between our model and the PCN model, where the
entanglements are not fixed in space, we consider the obtained results are reasonable.
%
%

In order to compare the computation cost (efficiency) of our model,
we report CPU times for several different simulation models on a PC
(Intel Xeon X5570 2.93GHz, Cent OS 4).
We chose a melt of Kremer-Grest chains with bead number $N=200$ as a reference, 
and made calculations by the present model and by the PCN model for
similar systems. Each simulation was performed for time steps
comparable to the longest relaxation time.
All the simulations were performed by serial simulation codes.
The Kremer-Grest simulation was made by COGNAC ver
7.1.1\cite{OCTA}. The other simulations were performed by our home-made
simulation codes.
For a Kremer-Grest simulation with the chain number $M=6$, the required time steps is about
$100,000$ and the CPU time was $60$ hours.
For our model, a simulation with $N=64$, $M=32$, $N_e=4$ and
$\zeta_s=1.0$,
required the $3000$ time steps and the CPU time was $30$ minutes.
For a PCN simulation with the average segment number of segments per
chain $Z=15$ and the number of chains $M=341$ required $200$ time steps
for the relaxation and the CPU time was $2$ minutes.
From these results, we can conclude that the computational cost
(efficiency) of our model is in between the Kremer-Grest model and the
PCN model. Judging from the coarse-graining level of these models, this
result is reasonable.

\section{Conclusion}

In this study, we proposed the multi-chain slip-spring model where the bead-spring chains are dispersed in space
 and connected by slip-springs. 
The entanglement effect is mimicked by the slip-springs and not by the
 hard-core (excluded volume) interaction between beads.
 Our model is located in the niche of the modeling of entangled polymer dynamics between conventional multi-chain simulations and single chain models. 
In our model, the set of state variables are the position of beads and the
 connectivity (bead indices) of the slip-springs.
Differently from the primitive chain network model (that is the other class of multi-chain model), 
the total free energy of the system is well-defined, and kinetic
 equations are designed based on the free energy and the detailed
 balance condition.
The free energy includes the repulsive interaction between beads, which compensate the
 attractive interaction artificially generated by the slip-springs. 
The explicit expression of linear relaxation modulus was also derived by the linear response theory.
{ A possible numerical scheme was proposed and simulations reproduced expected bead number dependence 
  in transitional regime between Rouse and entangled dynamics for the chain structure,
 the central bead diffusion and the linear relaxation modulus.}

In this model there are a couple of parameters ($N_s$, $\zeta_s$,
$\rho_0$ and $\phi$) of which effect on chain dynamics is not well understood. 
It would be an interesting work to explore the effects of these
parameters by theoretical methods as well as simulations.
We are performing simulations to scan these parameters and compare the resulting chain dynamics with the other models and experiments.
Another interesting topic is on the extension of this model by using its multi-chain nature, as reported for PCN model, to branch polymers, blends and copolymers, etc. The results for these attempts will be reported elsewhere.

\section*{ACKNOWLEDGMENTS}
TU is supported by Grant-in-Aid for Young Scientists B 22740273.
YM is supported by Grant-in-Aid for Scientific Research 23350113.
The authors thank Prof. A. E. Likhtman for useful comments and discussions.

\listoffigures

%
\newpage
\newpage
\end{document}